\documentclass[twocolumn,trackchanges]{aastex7}
\usepackage{amsmath}
\usepackage{amssymb}
\usepackage{graphicx}
\usepackage{booktabs}
\usepackage{overpic}
\usepackage{xcolor}

\begin{document}

\title{WFST Follow-up of S251112cm: Searching for an Optical Counterpart to a Subsolar-mass Compact-binary Merger Candidate}


\correspondingauthor{Wen Zhao, Zhiping Jin, Zigao Dai}
\email{wzhao7@ustc.edu.cn, jin@pmo.ac.cn, daizg@ustc.edu.cn}

\author[0000-0002-2242-1514,gname=Zhengyan,sname=Liu]{Zhengyan Liu}
\affiliation{Department of Astronomy, University of Science and Technology of China, Hefei 230026, China}
\affiliation{School of Astronomy and Space Sciences, University of Science and Technology of China, Hefei 230026, China}
\email{ustclzy@mail.ustc.edu.cn}

\author{Zelin Xu}
\affiliation{Department of Astronomy, University of Science and Technology of China, Hefei 230026, China}
\affiliation{School of Astronomy and Space Sciences, University of Science and Technology of China, Hefei 230026, China}
\email{sa21022025@mail.ustc.edu.cn}

\author[0000-0002-9092-0593]{Ji-an Jiang}
\affiliation{Department of Astronomy, University of Science and Technology of China,
Hefei 230026, China}
\affiliation{School of Astronomy and Space Sciences,
University of Science and Technology of China,
Hefei 230026, China}
\affiliation{National Astronomical Observatory of Japan,
National Institutes of Natural Sciences,
Tokyo 181-8588, Japan}
\email{jian.jiang@ustc.edu.cn}

\author[0000-0002-1330-2329]{Wen Zhao}
\affiliation{Department of Astronomy, University of Science and Technology of China, Hefei 230026, China}
\affiliation{School of Astronomy and Space Sciences, University of Science and Technology of China, Hefei 230026, China}
\affiliation{College of Physics, Guizhou University, Guiyang 550025, China}
\email{wzhao7@ustc.edu.cn}

\author[0000-0003-4977-9724]{Zhiping Jin}
\affiliation{Purple Mountain Observatory, Chinese Academy of Sciences, Nanjing 210023, China}
\email{jin@pmo.ac.cn}

\author[0000-0002-7835-8585]{Zigao Dai}
\affiliation{Department of Astronomy, University of Science and Technology of China, Hefei 230026, China}
\affiliation{School of Astronomy and Space Sciences, University of Science and Technology of China, Hefei 230026, China}
\email{daizg@ustc.edu.cn}

\author{Dezheng Meng}
\affiliation{Department of Astronomy, University of Science and Technology of China, Hefei 230026, China}
\affiliation{School of Astronomy and Space Sciences, University of Science and Technology of China, Hefei 230026, China}
\email{dezhengmeng@mail.ustc.edu.cn}

\author[0000-0002-7330-4756]{Yefei Yuan}
\affiliation{Department of Astronomy, University of Science and Technology of China, Hefei 230026, China}
\affiliation{School of Astronomy and Space Sciences, University of Science and Technology of China, Hefei 230026, China}
\email{yfyuan@ustc.edu.cn}

\author[0000-0002-6299-1263]{Xuefeng Wu}
\affiliation{Purple Mountain Observatory, Chinese Academy of Sciences,
Nanjing 210023, China}
\affiliation{School of Astronomy and Space Sciences,
University of Science and Technology of China,
Hefei 230026, China}
\email{xfwu@pmo.ac.cn}

\author[0000-0003-4200-4432]{Lulu Fan}
\affiliation{Department of Astronomy, University of Science and Technology of China,
Hefei 230026, China}
\affiliation{School of Astronomy and Space Sciences,
University of Science and Technology of China,
Hefei 230026, China}
\affiliation{Institute of Deep Space Sciences,
Deep Space Exploration Laboratory,
Hefei 230026, China}
\affiliation{College of Physics, Guizhou University,
Guiyang 550025, China}
\email{llfan@ustc.edu.cn}

\author[0000-0002-7660-2273]{Xu Kong}
\affiliation{Department of Astronomy, University of Science and Technology of China,
Hefei 230026, China}
\affiliation{School of Astronomy and Space Sciences,
University of Science and Technology of China,
Hefei 230026, China}
\affiliation{Institute of Deep Space Sciences,
Deep Space Exploration Laboratory,
Hefei 230026, China}
\email{xkong@ustc.edu.cn}

\author[0000-0002-9364-5419]{Xander J. Hall}
\affiliation{McWilliams Center for Cosmology and Astrophysics,
    Department of Physics,
    Carnegie Mellon University,
    5000 Forbes Avenue, Pittsburgh, PA 15213, USA}
\email{xhall@cmu.edu}

\author[0000-0002-6011-0530]{Antonella Palmese}
\affiliation{McWilliams Center for Cosmology and Astrophysics,
    Department of Physics,
    Carnegie Mellon University,
    5000 Forbes Avenue, Pittsburgh, PA 15213, USA}
\email{palmese@cmu.edu}

\author{Feng Li}
\affiliation{State Key Laboratory of Particle Detection and Electronics,
University of Science and Technology of China,
Hefei 230026, China}
\email{phonelee@ustc.edu.cn}

\author{Ming Liang}
\affiliation{National Optical Astronomy Observatory
(NSF's National Optical-Infrared Astronomy Research Laboratory),
950 N. Cherry Ave., Tucson, AZ 85726, USA}
\email{liangming@gmail.com}

\author{Binyang Liu}
\affiliation{Purple Mountain Observatory, Chinese Academy of Sciences,
Nanjing 210023, China}
\affiliation{School of Astronomy and Space Sciences,
University of Science and Technology of China,
Hefei 230026, China}
\email{byliu@pmo.ac.cn}

\author[0000-0002-3105-3821]{Zhen Wan}
\affiliation{Department of Astronomy, University of Science and Technology of China,
Hefei 230026, China}
\affiliation{School of Astronomy and Space Sciences,
University of Science and Technology of China,
Hefei 230026, China}
\email{zhen_wan@ustc.edu.cn}

\author[0000-0002-4372-0759]{Hairen Wang}
\affiliation{Purple Mountain Observatory, Chinese Academy of Sciences,
Nanjing 210023, China}
\affiliation{School of Astronomy and Space Sciences,
University of Science and Technology of China,
Hefei 230026, China}
\email{hairenwang@pmo.ac.cn}

\author[0000-0003-1617-2002]{Jian Wang}
\affiliation{Institute of Deep Space Sciences,
Deep Space Exploration Laboratory,
Hefei 230026, China}
\affiliation{State Key Laboratory of Particle Detection and Electronics,
University of Science and Technology of China,
Hefei 230026, China}
\email{wangjian@ustc.edu.cn}

\author[0000-0002-1517-6792]{Tinggui Wang}
\affiliation{Department of Astronomy, University of Science and Technology of China,
Hefei 230026, China}
\affiliation{School of Astronomy and Space Sciences,
University of Science and Technology of China,
Hefei 230026, China}
\email{twang@ustc.edu.cn}

\author[0000-0002-1463-9070]{Hongfei Zhang}
\affiliation{State Key Laboratory of Particle Detection and Electronics,
University of Science and Technology of China,
Hefei 230026, China}
\email{nghong@ustc.edu.cn}

\author[0000-0003-3728-9912]{Xianzhong Zheng}
\affiliation{Tsung-Dao Lee Institute and Key Laboratory for Particle Physics,
Astrophysics and Cosmology, Ministry of Education,
Shanghai Jiao Tong University,
Shanghai 201210, China}
\email{xzzheng@sjtu.edu.cn}

\author[0000-0003-0694-8946]{Qingfeng Zhu}
\affiliation{Department of Astronomy, University of Science and Technology of China,
Hefei 230026, China}
\affiliation{School of Astronomy and Space Sciences,
University of Science and Technology of China,
Hefei 230026, China}
\email{zhuqf@ustc.edu.cn}

\begin{abstract}
    Electromagnetic counterparts to gravitational-wave (GW) sources probe the properties, environments, and evolution of compact-object mergers. S251112cm belongs to an emerging class of GW candidates with possible subsolar-mass components. We present an optical counterpart search for S251112cm with the Wide Field Survey Telescope (WFST). The observations began 20.2 hr after the GW trigger and continued for three nights, covering approximately $780~\mathrm{deg}^{2}$ and $51\%$ of the localization probability in the updated skymap. We searched for newly emerging, rapidly evolving, off-nuclear optical transients and identified four candidates whose host-galaxy distances are broadly consistent with the GW distance estimate. However, all four evolve substantially more slowly than AT~2017gfo, ruling out an AT~2017gfo-like origin and disfavoring their association with S251112cm as rapidly evolving kilonova counterparts. Combining WFST and DECam observations increases the covered localization probability to approximately \(67\%\). Conditional on the counterpart lying within this footprint, we constrain a grid of binary-neutron-star kilonova models. For viewing angles $\theta_{\rm obs}<60^{\circ}$, with the other parameters fixed to the best-fitting values for AT~2017gfo, models with dynamical ejecta mass $M_{\rm dyn}\gtrsim0.01\,M_{\odot}$ or wind ejecta mass $M_{\rm wind}\gtrsim0.05\,M_{\odot}$ are disfavored over most of the covered GW probability. Beyond kilonova models, our phenomenological analysis constrains rapidly evolving optical transients with \(T_{\rm rise,1/2}\leq2\) days to peak absolute magnitudes fainter than approximately \(-13\) mag. To our knowledge, WFST provides the strongest optical constraints to date in the northern sky on possible rapidly evolving counterparts to this new class of GW candidates involving subsolar-mass compact objects. The comparable depths and complementary sky coverage of WFST and DECam extend these constraints across a larger fraction of the localization region at similar sensitivity.
\end{abstract}
\keywords{\uat{Gravitational wave astronomy}{675} --- \uat{Compact objects}{288} --- \uat{Optical observation}{1169}}

\section{Introduction} 
    The first direct detection of gravitational waves (GWs) from the binary black hole (BBH) merger GW150914 opened a new window for studying compact objects and their coalescences \citep{Abbott_2016_GW150914}. The subsequent detection of the binary neutron star (BNS) merger GW170817, together with its electromagnetic (EM) counterparts including the short gamma-ray burst (sGRB)~170817A, a multi-wavelength afterglow extending from X-rays to radio wavelengths, and the kilonova (KN) AT~2017gfo, marked the beginning of GW multi-messenger astronomy \citep{Abbott_2017_GW170817,Abbott_2017_Multimessenger,Troja_2017,coulter_swope_2017,pian_spectroscopic_2017,DAvanzo_2018}. The EM counterpart observations, combined with GW data, can be applied to a wide range of scientific studies, such as the origin of heavy elements, the neutron-star equation of state, and constraints on the Hubble constant \citep{Kasen_2017,Abbott_2017Natur,Coughlin_2019b}.

    Subsolar-mass compact binaries are commonly defined as systems containing at least one component with a mass below $1\,M_{\odot}$. Compact objects in this mass range are not expected to form through conventional stellar evolution. In particular, stellar-origin black holes are expected to be more massive than the Sun, while the formation of extremely low-mass neutron stars through ordinary core-collapse supernovae (CCSNe) is difficult. Subsolar-mass compact objects may instead originate from primordial black holes, nonstandard neutron-star formation channels, strange stars, or other exotic compact objects \citep{Crescimbeni_2024,Metzger_2024}. Dedicated searches have been performed in GW data from the first three observing runs of LIGO and Virgo, covering both comparable-mass binaries and systems with extreme mass ratios, but no statistically significant signal was identified \citep{Abbott_2018_SSM,Abbott_2019_SSM,Abbott_2022_SSM,Nitz_2022}. More recent searches have extended the parameter space to systems with large tidal deformabilities, as expected for extremely low-mass neutron stars \citep{Kacanja_2026}. Nevertheless, the component masses inferred from a GW signal may not uniquely determine the nature of a subsolar-mass compact object, motivating complementary searches for associated EM emission.

    The EM signatures of subsolar-mass compact object mergers are expected to depend strongly on the nature and formation environment of the binary. A binary composed of two black holes is generally expected to merge without detectable EM emission in vacuum, although radiation may be produced if the merger occurs in a dense gaseous environment, such as an active galactic nucleus (AGN) disk \citep{Graham_2020,Chen_2024c}. In contrast, mergers involving subsolar-mass neutron stars may eject substantial amounts of matter, potentially powering kilonova emission \citep{Corman_2026}. Mergers of neutron stars with subsolar-mass black holes can also produce kilonova emission \citep{Markin_2023}. Subsolar-mass strange-star mergers could generate ejecta with a different composition and radiative signature \citep{Gao_2026}. Another proposed formation channel involves fragmentation or fission during the collapse of a rapidly rotating massive star. In this scenario, subsolar-mass neutron stars may form and merge within a collapsar disk or an expanding stellar envelope, producing a kilonova embedded within a supernova (SN), sometimes referred to as a ``superkilonova'' \citep{Metzger_2024,Kasliwal_2025}. However, these predictions remain highly uncertain because the ejecta properties depend on the poorly constrained nature, equation of state, mass ratio, and formation environment of the compact objects. The luminosity, color, and evolution timescale of the associated transient may therefore differ substantially from those of AT~2017gfo, making it more difficult to distinguish the potential counterpart from unrelated transients. 

    During the fourth observing run (O4), S250818k was reported by the LIGO--Virgo--KAGRA Collaboration (LVK) as a low-significance compact binary merger candidate with a possible subsolar-mass component \citep{2025GCN.41437....1L}. S250818k has a relatively high false alarm rate (FAR) of one per 2.1 years. Several studies have discussed a potential association between S250818k and the spatially and temporally coincident Type~IIb SN~2025ulz, although the physical association remains inconclusive \citep{Kasliwal_2025,hallAT2025ulzS250818kLeveraging2026}. More recently, on 2025 November 12, LVK reported S251112cm, another compact-binary merger candidate containing at least one possible subsolar-mass component \citep{2025GCN.42650....1L,2025GCN.42690....1L}. S251112cm has a lower FAR of approximately one per four years, making it a particularly important target within this emerging class of GW candidates. Identifying an associated EM counterpart, or placing meaningful limits on its emission, could help distinguish between the possible natures and formation channels of subsolar-mass compact objects. Its relatively nearby distance ($\sim100~{\rm Mpc}$) therefore motivated rapid, wide-field optical follow-up campaigns, including a substantial effort by the Vera C. Rubin Observatory (Rubin-LSST), to search for possible counterparts and constrain their observational properties \citep{2025GCN.42677....1A, 2025GCN.42691....1H, 2025GCN.42707....1M, 2025GCN.43257....1A,Vieira_2026,Hall_2026}.

    In this work, we present the follow-up campaign conducted by the 2.5-meter Wide Field Survey Telescope (WFST) for S251112cm. In Section \ref{sec_2}, we introduce the GW event S251112cm and our WFST follow-up observations. The reduction procedures for the follow-up data are presented in Section \ref{sec_3}. In Section \ref{sec_4}, we describe the scope and results of the EM counterpart search. We then discuss whether the identified candidates could be associated with the GW event. In Section \ref{sec_5}, we use the non-detection to constrain the optical counterpart of S251112cm. In Section \ref{sec_6}, possible EM-counterpart scenarios beyond our search scope are discussed. Finally, we summarize our search campaign and constraint results in Section \ref{sec_7}. Throughout this study, we adopt a standard $\Lambda$CDM cosmology with parameters $H_0=67.7\,\text{km s}^{-1}\,\text{Mpc}^{-1}$, $\Omega_M=0.31$ and $\Omega_\Lambda=0.69$ \citep{Planck2018}.
    
\section{S251112cm and WFST follow-up campaign} \label{sec_2}
    The GW event S251112cm was detected by the LIGO Livingston, LIGO Hanford, and Virgo detectors at 15:18:45.362 UTC on November 12, 2025 \citep{2025GCN.42650....1L}. Based on the subsequent updated alert \citep{2025GCN.42690....1L}, S251112cm is a compact object merger candidate with a FAR of 1 per $\sim4$ years, characterized by at least one subsolar-mass component. Its luminosity distance was estimated to be $93\pm27$ Mpc with a 90\% sky localization area of $1681\,\text{deg}^2$. Given its proximity, S251112cm provides a valuable opportunity to search for and investigate the potential EM counterpart of a compact binary coalescence involving a subsolar-mass object.

    We conducted a follow-up campaign for S251112cm using WFST \citep{2025GCN.42722....1L}, which is installed at the summit of Saishiteng Mountain near Lenghu, Qinghai province, China \citep{Deng_2021}. Featuring a 2.5-meter primary mirror and a 6.55 deg$^{2}$ field of view (FoV), WFST is one of the most powerful transient survey facilities \citep{Wang_2023,conroy2023china}, with extragalactic transients like GW EM counterparts, SNe, and tidal disruption events serving as its primary targets. WFST achieved first light in September 2023, a few months after the start of O4. To search for potential EM counterparts of GW events detected in O4, an optimized follow-up strategy for WFST target of opportunity (ToO) observations was developed in \cite{Liu_2023}. This ToO strategy was subsequently applied to observations of GW events S240422ed and S250206dm \citep{Liu_2026a}.

\begin{figure*}[htbp]
        \centering
        \begin{overpic}[width=0.98\textwidth]{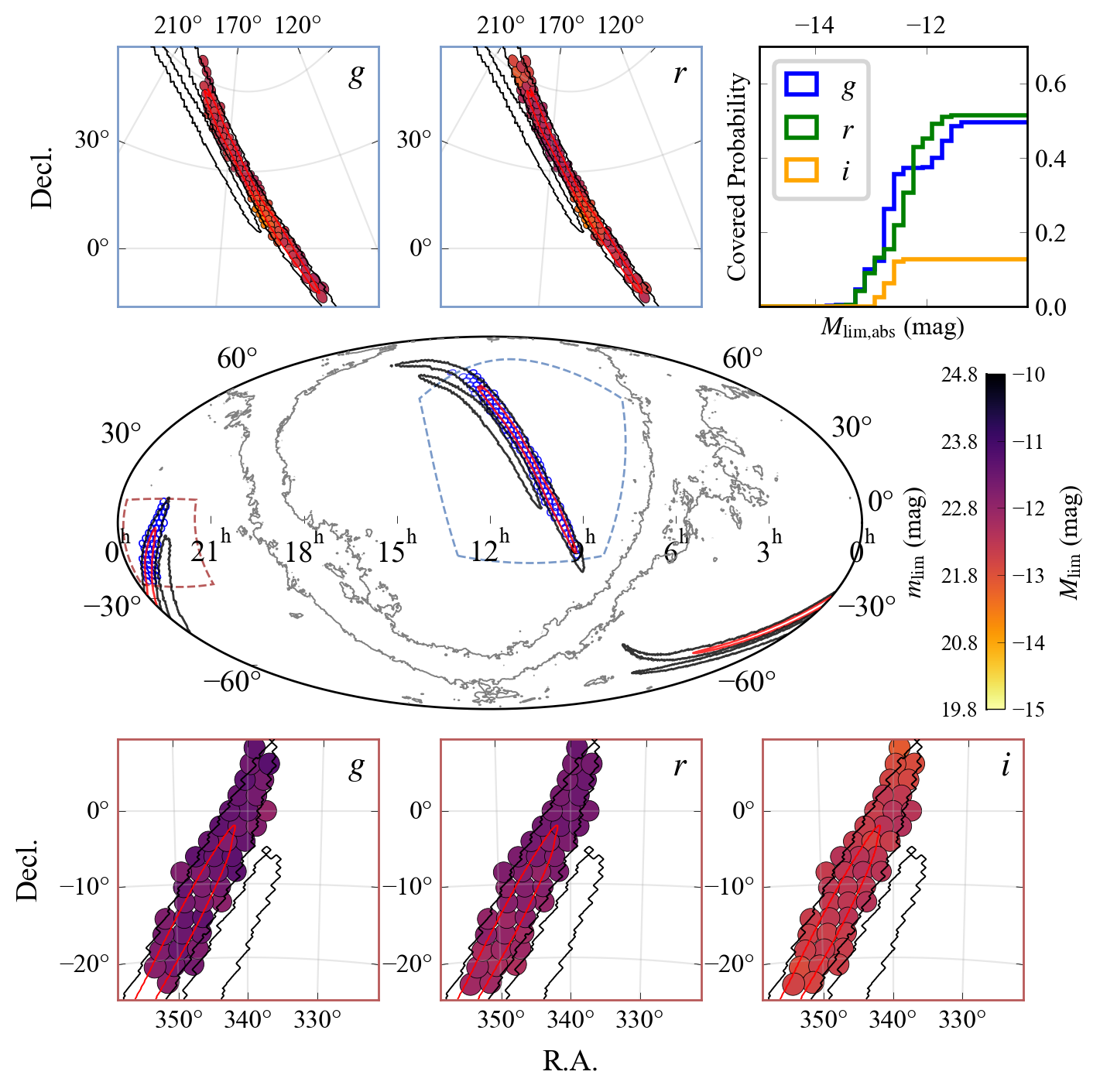}
        \end{overpic}
        \caption{Skymap coverage and multi-band depths of WFST on the first night for S251112cm. Middle: The overall coverage by WFST, overlaid with the Bilby skymap of S251112cm from \cite{2025GCN.42690....1L}, showing the 50\% (red) and 90\% (black) probability contours. Galactic extinction is also indicated, with the contour of $E(B-V)=0.3$ shown for reference \citep{Schlafly_2011}. A WFST pointing is represented by a circle with a radius of $1.4^\circ$. Top: Zoomed-in view of the covered regions in the $g$ and $r$ bands over the R.A. range from 9\textsuperscript{h} to 13\textsuperscript{h}, corresponding to the region outlined by the blue dashed line in the middle panel. The color map represents the 5$\sigma$ limiting magnitudes $m_{\rm lim}$ and limiting absolute magnitudes $M_{\rm lim}$, calculated assuming the mean GW luminosity distance of 93 Mpc. The top-right panel shows the cumulative GW localization probability covered by WFST as a function of limiting absolute magnitudes in the $g$, $r$, and $i$ bands. Bottom: Zoomed-in view of the covered regions in $g$, $r$, and $i$ bands over the R.A. range from 22\textsuperscript{h} to 24\textsuperscript{h}, corresponding to the region outlined by the red dashed line.}
        \label{fig:coverage}
    \end{figure*}

    For S251112cm, WFST observations started approximately 20.2 hours (11:30 UTC on November 13, 2025) after the GW detection and lasted for three days. The overall coverage by WFST and limiting magnitudes on the first night are shown in Figure \ref{fig:coverage}, based on the updated skymap. The follow-up plan was based on the initial skymap \citep{2025GCN.42650....1L} and primarily focused on the northern part of the skymap. A filter combination of $g$, $r$, and $i$ bands was adopted, with exposure times ranging from 45 s to 90 s. To evaluate the search capability, in Figure \ref{fig:coverage}, the limiting magnitudes in each band are converted to absolute magnitudes using the mean GW luminosity distance of 93 Mpc. Galactic foreground extinction is corrected using the $E(B-V)$ map from \cite{Schlafly_2011} with $R_V = 3.1$, while host extinction was not considered. Our follow-up observations are sensitive enough to detect an AT~2017gfo-like kilonova, whose peak absolute magnitudes are $\sim-16$ mag in the optical bands \citep{Villar_2017a}.

    The observing strategies for the second and third nights were the same as that of the first night. Although the localization of S251112cm was updated $\sim$2.4 days after the GW detection, the subsequent follow-up plan remained unchanged to ensure continuity of observations. Due to the enlargement of the 90\% sky area from $1220\,\text{deg}^2$ to $1681\,\text{deg}^2$ after the update, part of the northern region was not covered, as shown in Figure \ref{fig:coverage}. In total, $\sim51\%$ ($\sim65\%$) of the localization probability was covered by the three-night WFST follow-up campaign for the updated skymap (the initial skymap), corresponding to an area of $\sim780\,\text{deg}^2$.

    In addition to the ToO follow-up campaign, part of the northern skymap overlapped with the WFST Wide-field Survey (WFS; \cite{Wang_2023}). WFS is a key program planned for the WFST 6-year survey. WFS covers an area of $\sim8000\,\text{deg}^2$ in the northern sky in $u$, $g$, and $r$ bands with a single exposure of 30 s, leading to about 90 visits per pointing in each band over 6 years. The earliest WFS observations started $\sim6.4$ hours post-merger, covering $\sim6.5\%$ of the updated skymap on that night. 

\section{Data Reduction} \label{sec_3}
    The raw images from the follow-up observations of S251112cm were processed using the WFST pipeline, which is primarily based on the Rubin-LSST image processing pipeline (hereafter the ``LSST pipeline''; \citealp{boschOverviewLSSTImage2018}). Designed as a modular software suite for processing massive survey data, the LSST pipeline encompasses key steps including data ingestion, instrument signature removal, point spread function (PSF) modeling, astrometric and photometric calibration, source detection and deblending, image alignment, image subtraction, and alert product generation. The WFST pipeline was modified from the LSST pipeline in two main aspects: the addition of extension files defining the CCD configuration and the replacement of the subtraction software with Saccadic Fast Fourier Transform (\texttt{SFFT}; \citealp{Hu_2022}) to optimize and expedite the subtraction process. Details of the WFST pipeline are available in \cite{Cai_2025}.

    For the region overlapping with the WFS of WFST, the EM counterpart search was conducted promptly owing to the availability of archival reference images \citep{2025GCN.42722....1L}. For the region with no archival images, template images were subsequently acquired from 2025-12-05 to 2026-02-15 to ensure that any potential counterpart had faded sufficiently. In total, millions of alerts were generated through basic image reduction and subtraction using the WFST pipeline. We screened these alerts to exclude contaminants using the following filtering criteria:
    \begin{itemize}
    \item[1.] Artifact rejection: 
    We require at least two detections ($N_{\rm det}\geq2$), with at least one passing the real-bogus (RB) classifier ($N_{\rm real}\geq1$; \citealp{Liu_2025}).
    \item[2.] Variable star exclusion: Alerts located close to known point sources are excluded.
    \item[3.] Bright source proximity: Alerts around bright sources ($m < 15$ mag) are removed to avoid false detections caused by saturation.
    \item[4.] Moving object elimination: Alerts corresponding to known Solar System objects are excluded.
    \end{itemize}

    The number of alerts remaining after each filtering step is shown in Table \ref{tab:alert_filter}. Crossmatching with variable stars, bright sources, and galaxy centers was performed using the Pan-STARRS1 (PS1) DR2 and Gaia DR3 catalogs \citep{Chambers_2016,Magnier_2020,GaiaCollaboration_2023}. Point sources and extended sources are distinguished based on the star classification probability for the Gaia database and the difference between PSF and Kron photometry for the PS1 database (with sources satisfying $m_{i,\text{PSF}} - m_{i,\text{Kron}} < 0.05$ classified as stars\footnote{\href{https://outerspace.stsci.edu/display/PANSTARRS/How+to+separate+stars+and+galaxies}{https://outerspace.stsci.edu/display/PANSTARRS/
    How+to+separate+stars+and+galaxies}}). To exclude moving objects, the remaining alerts were crossmatched with the Sky Body Tracker to rule out Solar System objects \citep{SkyBot_2006}. Finally, 6,413 alerts remained after applying this inclusive set of filters to avoid missing potential candidates. 

    \begin{table}[htbp]
        \centering
        \footnotesize
        \renewcommand{\arraystretch}{1.2}
        \setlength{\tabcolsep}{0.1pt}
        \caption{\noindent\textbf{Numbers of alerts remaining after successive filtering and crossmatching steps.}} 
        \begin{tabular}{l*{2}{c}}
            \toprule 
            Description           & Filter                      & \# \\
            \midrule
            Total                 &                             & 3,802,280           \\
            Multiple detections   & $N_\text{det}\geq2$         & 755,543           \\
            Real source           & $N_\text{real}\geq1$        & 29,849             \\
            Not a variable star     & $d_\text{point}\geq1''$     & 9,496              \\
    Far from bright sources        & $d_\text{15mag}\geq3'', d_\text{12mag}\geq30''$ & 8,562 \\
            Not a moving object     & $d_\text{SkyBot}\geq3''$    & 6,413               \\
            \midrule
            \multicolumn{3}{l}{Visual inspection} \\
            \midrule
            Off-nuclear transients & $d_\text{host}>0.5''$ & 325 \\
            No detection pre-merger &  & 174 \\
            
            \midrule
            \multicolumn{3}{l}{Host galaxy crossmatch} \\
            \midrule
            Host matched & & 164 \\
            Spectroscopic redshift available & & 53\\
            \bottomrule
        \end{tabular}
        \label{tab:alert_filter}
    \end{table}

\section{Counterpart Search \& Analysis} \label{sec_4}
    \subsection{Scope of the Counterpart Search}
        The EM emission associated with a subsolar-mass compact binary merger remains highly uncertain. In this work, we restrict our search to off-nuclear optical transients that emerge after the merger and evolve on timescales of hours to several days. This choice is motivated by the cadence and duration of our follow-up observations and encompasses kilonova-like emission as well as other rapidly evolving optical transients. Given the short temporal baseline of our observations, variability spatially coincident with a galaxy nucleus cannot be reliably distinguished from intrinsic AGN variability or image-subtraction residuals. We therefore focus on off-nuclear sources and do not require the candidates to follow a specific kilonova model or color evolution.

        Other possible counterparts, including emission from binary black hole mergers in AGN disks and superkilonova-like events, are not included in the present search. Emission from the former is likely to have a low contrast relative to intrinsic AGN variability in the subsolar-mass regime, whereas a systematic search for the latter requires sufficiently deep pre-merger imaging and longer-term monitoring. We discuss these possibilities and the corresponding limitations of our search in Section~\ref{sec_6}.
        
        Within this scope, we search for newly emerging, rapidly evolving, off-nuclear optical transients following the GW trigger. Consistency between the distance of a candidate host galaxy and the GW luminosity-distance posterior is adopted as the primary astrophysical criterion for assessing a possible association. 
 
    \subsection{Candidate Selection}
       \begin{figure*}[htbp]
            \centering
            \begin{overpic}[width=0.9\textwidth]{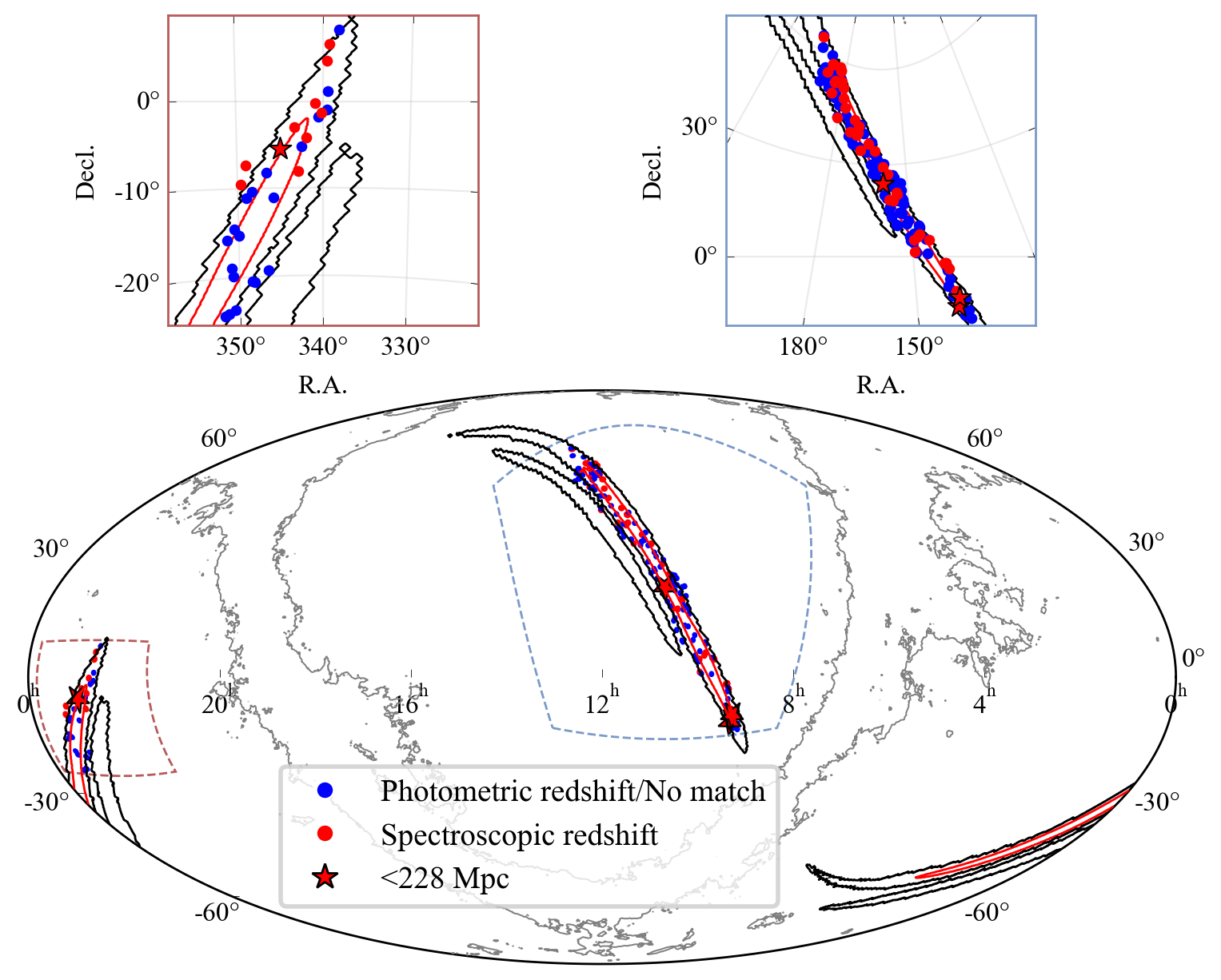}
            \end{overpic}
            \caption{Sky distribution of off-nuclear extragalactic transients without pre-merger detections. Blue circles indicate sources with photometric redshifts or no host match, and red circles indicate sources with spectroscopic redshifts. Red stars highlight the four candidates with estimated luminosity distances below 228 Mpc. The upper panels show enlarged views of the regions outlined by dashed lines in the lower panel.}
            \label{fig:candidate_loc}
        \end{figure*}
        
        To construct a candidate sample within the search scope defined above, we visually inspected the filtered alerts to further exclude artifacts and other contaminants. In total, 325 alerts were identified as off-nuclear extragalactic transient candidates, which are summarized in Table~\ref{tab:alert_filter}. To exclude transients that occurred before the GW detection, forced photometry was performed on WFST archival images taken before the merger. In addition, these transients were crossmatched with public survey databases, including the Transient Name Server \citep[TNS;][]{TNS}, the ALeRCE explorer \citep{Forster_2021} for data of the Zwicky Transient Facility \citep[ZTF;][]{Bellm_2019,graham_zwicky_2019}, and the Asteroid Terrestrial-impact Last Alert System \citep[ATLAS;][]{Tonry_2018,Smith_2020} forced photometry server \citep{ATLAS_force}. Ultimately, 174 transients remained with no pre-merger detections. To obtain their host redshifts, these transients were sequentially crossmatched with catalogs including the Sloan Digital Sky Survey (SDSS) Data Release 17 spectroscopic catalog \citep{Abdurrouf_2022}, the Dark Energy Spectroscopic Instrument (DESI) Data Release 1 \citep{desicollaboration2025datarelease1dark}, the GLADE+ catalog \citep{Dalya_2022} (spectroscopic and photometric redshifts), the WISE-PS1-STRM photometric redshift catalog \citep{WISE-PS1-STRM}, the photometric redshift catalog of the Legacy Survey Data Release 11 \citep{Zhou_2023}, and the PS1-STRM photometric redshift catalog \citep{Beck_2021}. The crossmatch radius ranged from 0.5$^{\prime\prime}$ to 60$^{\prime\prime}$, where the adopted matching radius increased with decreasing galaxy distance.
    
        \begin{figure}[htbp]
            \centering
            \begin{overpic}[width=0.472\textwidth]{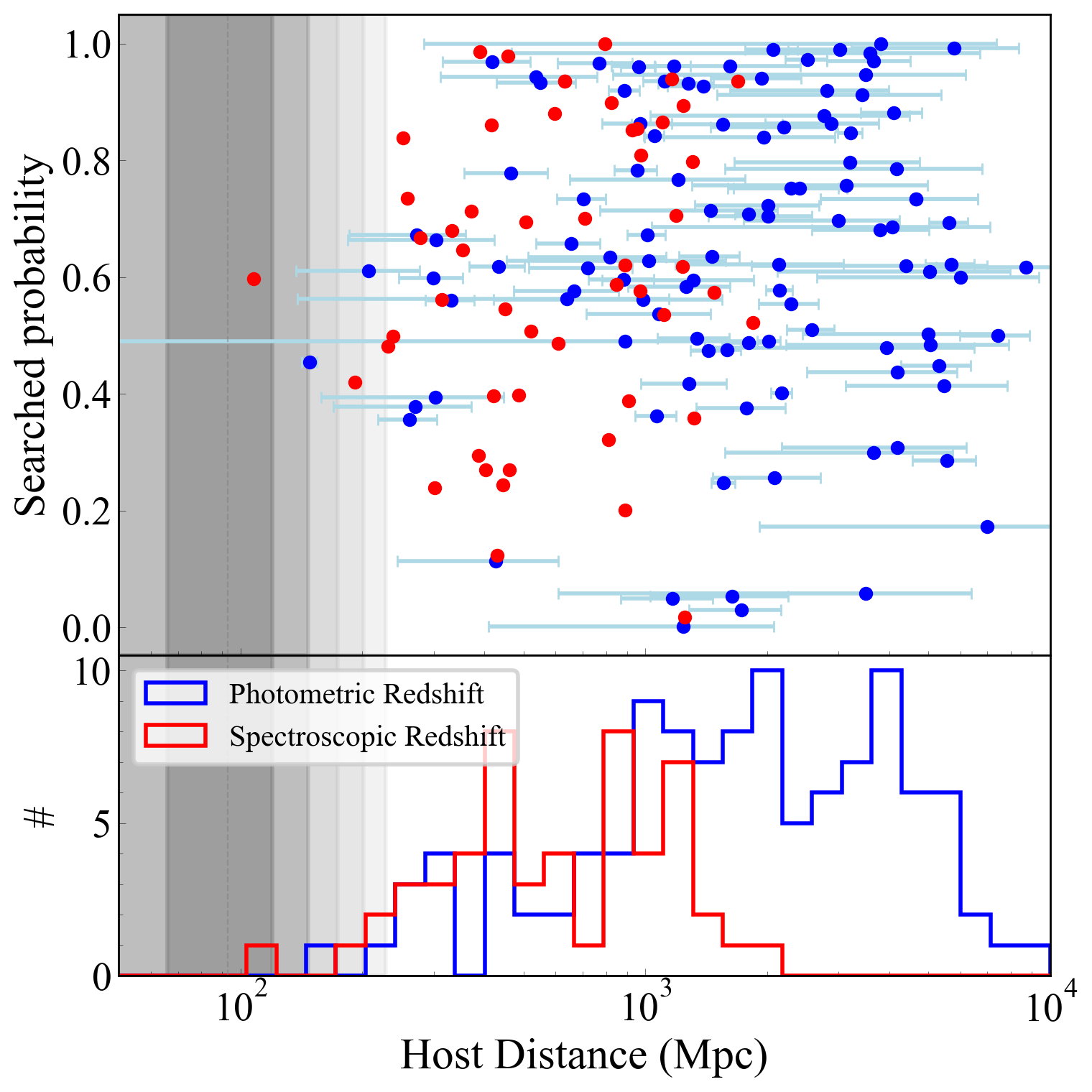}
            \end{overpic}
            \caption{Distributions of host distances and sky locations for off-nuclear extragalactic transients without pre-merger detection. The searched probability at each candidate position is the cumulative posterior probability of all skymap pixels with probability densities equal to or higher than that of the pixel containing the candidate, calculated using \texttt{ligo.skymap} \citep{Singer_2016_a, Singer_2016_b}. Smaller values indicate that the candidate lies in a region more strongly favored by the GW sky localization. Based on the GW distance estimate, the shaded gray regions represent the mean estimated distance (93 Mpc) $\pm1\sigma$, $\pm2\sigma$, ..., $\pm5\sigma$, where $\sigma$ is the distance uncertainty (27 Mpc). The transparency decreases with increasing $\sigma$-multiple to visually emphasize the central region. Distance errors for the photometric redshift sample are derived by sampling the redshift posterior.}
            \label{fig:dist_distri}
        \end{figure}
    
        \begin{figure*}[htbp]
            \centering
            \begin{overpic}[width=0.94\textwidth]{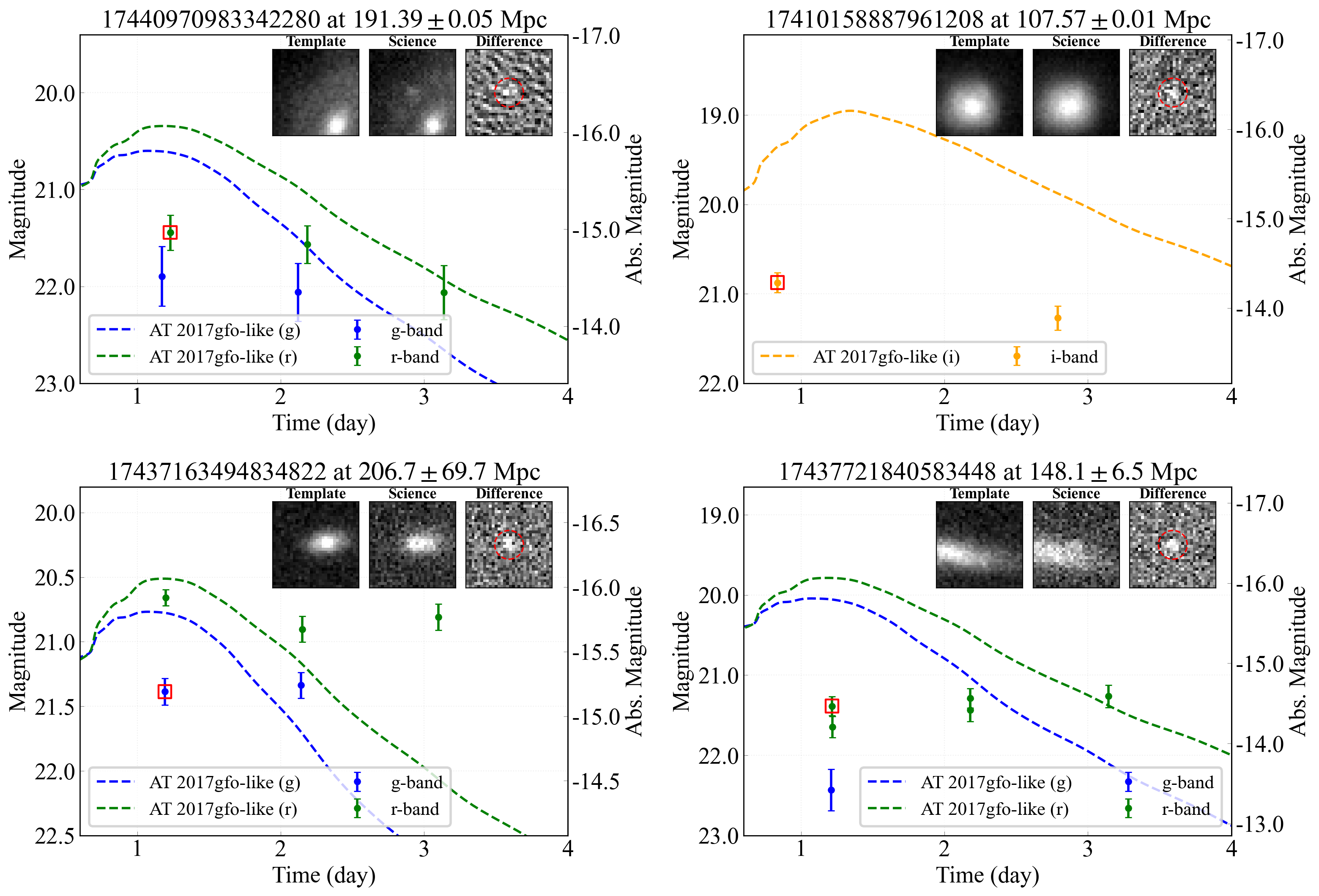}
            \end{overpic}
            \caption{Light curves of the four candidates located within $\pm5\sigma$ of the mean estimated distance. The luminosity distances of both the candidates and the AT~2017gfo-like kilonova are determined from their spectroscopic or photometric redshifts. The adopted distance for each object is indicated in the corresponding panel title, with uncertainties shown for distances derived from spectroscopic (upper panels) or photometric redshifts (lower panels). The template, science, and difference-image cutouts corresponding to the first detection are shown in the upper-right corner of each panel. The candidate position is marked by a red dashed circle in each cutout, while the corresponding photometric measurement in the light curve is marked by a red open square.}
            \label{fig:LCs}
        \end{figure*}

        \begin{table*}
            \centering
            \caption{Information on the four candidates located within $\pm5\sigma$ of the mean estimated distance. The distance uncertainties account only for redshift measurement errors.}
            \begin{tabular}{ccccc}
            \hline
            ID & R.A. (deg) & Decl. (deg) &
            $D_{\rm L}$ (Mpc) & Detection bands \\
            \hline
            17440970983342280 & 159.26312 & 23.19631 & $191.39 \pm 0.05$ & $g,r$ \\
            17410158887961208 & 345.00268 & -5.45435 & $107.57 \pm 0.01$ & $i$ \\
            17437163494834822 & 139.36999 & -11.34425 & $206.7 \pm 69.7$ & $g,r$ \\
            17437721840583448 & 138.81689 & -9.56588 & $148.1 \pm 6.5$ & $g,r$ \\
            \hline
            \end{tabular}
            \label{tab:four_candidates}
        \end{table*}

        The crossmatch results are summarized in Table \ref{tab:alert_filter}. The sky distribution of 174 transients without pre-merger detection is shown in Figure \ref{fig:candidate_loc}, where blue circles indicate sources with photometric redshifts or no host match, while red circles indicate sources with spectroscopic redshifts. Their host distance distribution is plotted in Figure \ref{fig:dist_distri}, omitting sources without host identification or with large photometric redshift errors ($z_{\rm photo,err} \geq z_{\rm photo}$). Most transients fall outside the GW-inferred distance range, being located at distances significantly greater than $93\pm27$ Mpc. Adopting a conservative upper bound of 228 Mpc (mean distance + $5\sigma$), only four transients remain within a preliminary distance cut. 
        
    \subsection{Candidate Analysis}
        Multiband light curves and cutout images of the four transients are shown in Figure~\ref{fig:LCs}, where light curves of AT~2017gfo are also shown. The basic information on the four transients is also summarized in Table~\ref{tab:four_candidates}. To assess the four candidates within the class of rapidly evolving post-merger transients targeted by this work, we use the kilonova associated with a conventional BNS merger as an empirical reference \citep{Markin_2023,Corman_2026}. The ejecta mass, velocity, and composition of a subsolar-mass merger could differ substantially from those inferred for GW170817. Therefore, the comparison with AT~2017gfo should be regarded as a benchmark rather than a template encompassing all possible counterparts.

        Compared with AT~2017gfo, all four candidates are fainter and show substantially slower evolution over the observed epochs. Their low luminosities alone do not exclude a kilonova origin, since a subsolar-mass merger could produce a counterpart fainter than AT~2017gfo. However, under the hypothesis that any of these candidates is associated with S251112cm, the onset time is fixed by the GW merger time. None of the four candidates exhibits the rapid post-merger photometric evolution characteristic of AT~2017gfo. Their observed light curves therefore rule out an AT~2017gfo-like kilonova interpretation and strongly disfavor an association with S251112cm as a rapidly evolving kilonova counterpart. Although the lack of sufficiently deep pre-merger imaging prevents us from determining whether these sources were already active before the GW trigger, their slow post-merger evolution is more naturally explained by unrelated faint extragalactic transients, such as SN impostors \citep{Pastorello_2019a}, intrinsically faint SNe \citep{Das_2025a,Teja_2026}, or dust-obscured SNe \citep{Jencson_2019}.

\section{Model Constraints} \label{sec_5}
    Under the assumption that none of the identified candidates are genuine counterparts, the limiting magnitudes obtained from our observations of this nearby event place meaningful constraints on theoretical models of subsolar-mass mergers. Here, a KN model from BNS mergers and a phenomenological transient model are employed for analysis. 
    \subsection{Joint Analysis with DECam}
         To place more reliable constraints, we combined the WFST observations with follow-up data obtained using the Dark Energy Camera \citep[DECam;][]{flaugher_dark_2015,Hall_2026} to obtain joint constraints over a larger fraction of the GW localization region. \citet{Hall_2026} identified the Type~IIb SN~2025adtq, which was detected by WFST before the merger, and discussed its possible association with S251112cm in the context of the superkilonova scenario. The association, however, remains inconclusive. Apart from this source, no viable post-merger counterpart was identified. We therefore use the DECam nondetections, together with the WFST observations, to constrain post-merger optical counterparts.
    
        \begin{figure}[htbp]
        \centering
        \begin{overpic}[width=0.47\textwidth]{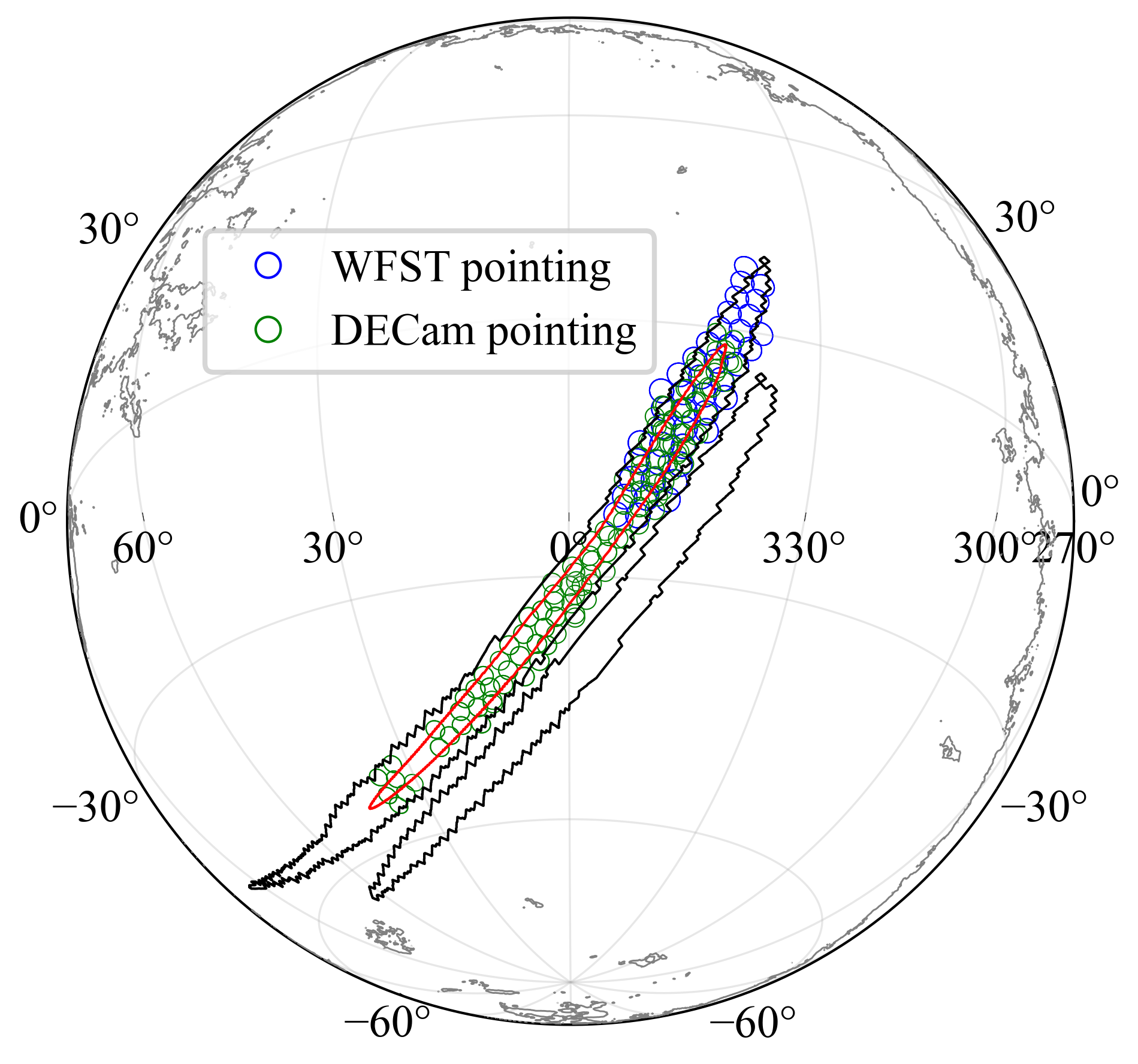}
        \end{overpic}
        \caption{Sky coverage of the WFST and DECam follow-up observations in the southern localization region, shown in a globe projection. Blue and green circles with radii of $1.4^\circ$ and $1.1^\circ$ approximate the WFST and DECam fields of view, respectively, illustrating their relative coverage and overlap. The 50\% and 90\% credible regions of the updated Bilby skymap of S251112cm \citep{2025GCN.42690....1L} are outlined in red and black, respectively. The grey contour marks Galactic extinction at $E(B-V)=0.3~\mathrm{mag}$ \citep{Schlafly_2011}.}
        \label{fig:DECam_WFST_coverage}
        \end{figure}
    
        The follow-up observations of DECam were conducted from $\sim2$ days to $\sim50$ days after the event, spanning a total of six epochs in $g$ and $i$ bands. The DECam follow-up observations covered the southern part of the GW localization region. The observations reach $5\sigma$ limiting magnitudes of 21--24 mag in the $g$ band and 21--23 mag in the $i$ band, closely matching the depths of the WFST observations. The comparable depths provide more uniform constraints across the surveyed region. For the specific schedule and details of the DECam observations, see \citet{Hall_2026}. The WFST and DECam coverage in the southern localization region is shown in Figure~\ref{fig:DECam_WFST_coverage}. The combined observations cover $\sim67\%$ ($\sim80\%$) of the localization probability in the updated (initial) skymap. Unless otherwise stated, the constraints presented below are conditional on the counterpart being located within the region covered by either WFST or DECam and therefore do not account for the possibility that it lies outside the observed footprint.
    
        \subsection{BNS Model}
        For the BNS-merger scenario, we use the KN radiative transfer code \texttt{POSSIS} \citep{Bulla_2019,Bulla_2023,Ahumada_2026}, which includes dynamical ejecta and post-merger disk wind ejecta. Tidal dynamical ejecta is concentrated near the equatorial plane, whereas shock-driven dynamical ejecta is preferentially ejected toward the poles. Disk winds generally have a broader angular distribution. We adopt the simulation grid results from \citet{Ahumada_2026}. The variable parameters and their discrete grid values are as follows: dynamical ejecta mass $M_{\rm dyn} \in [0.001, 0.005, 0.01, 0.02]\,M_\odot$, wind ejecta mass $M_{\rm wind} \in [0.01, 0.05, 0.09, 0.13]\,M_\odot$, dynamical ejecta velocity $v_{\rm dyn} \in [0.12, 0.15, 0.2, 0.25]\,c$, wind ejecta velocity $v_{\rm wind} \in [0.03, 0.05, 0.1, 0.15]\,c$, dynamical ejecta electron fraction $Y_{e{\rm ,dyn}} \in [0.15, 0.2, 0.25, 0.3]$, wind ejecta electron fraction $Y_{e,{\rm wind}} \in [0.2, 0.3, 0.4]$, and viewing angle, parameterized by $\cos\theta_{\rm obs}$, which is sampled uniformly from 0 to 1 in increments of 0.1. In total, the model grid contains 33,792 parameter combinations.
    
        To evaluate the overall constraining power of the joint follow-up observations, we use \texttt{dynesty} to draw nested samples from the covered portion of the three-dimensional GW localization posterior and evaluate the detectability of each model at every sampled sky position and distance. For each sampled position, the model light curves from all parameter combinations are compared against the observational limiting magnitudes at the corresponding positions to determine whether they are excluded. To align with standard operational search criteria, the exclusion condition requires at least two detections. Finally, the overall constraint is derived by averaging the exclusion fractions across all sampled positions.
    
        \begin{figure}[htbp]
        \centering
        \begin{overpic}[width=0.47\textwidth]{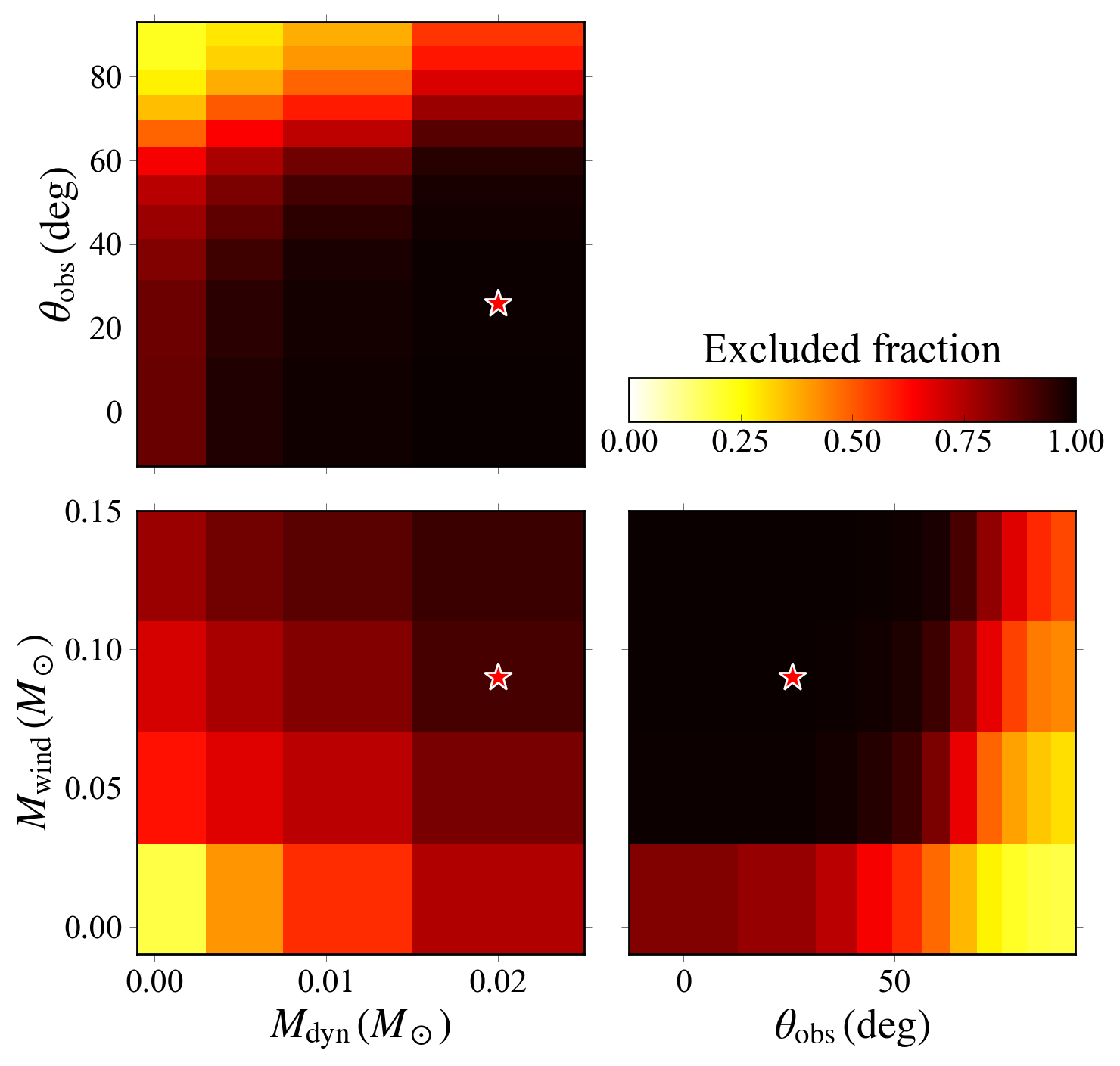}
        \end{overpic}
        \caption{Constraints on the ejecta masses and viewing angle, with other parameters ($v_{\rm ej}$, $Y_e$) fixed to the best-fit values of AT~2017gfo. The color scale indicates the grid excluded fraction of the parameter space. Red stars mark the inferred ejecta masses and viewing angle for AT~2017gfo, shown for comparison.}
        \label{fig:M_ej_constrain}
        \end{figure}

        Constraints on the dynamical and wind ejecta masses, as well as the viewing angle, are presented in Figure \ref{fig:M_ej_constrain}, where other model parameters ($v_{\rm ej}$, $Y_e$) are fixed based on the fitting results of AT~2017gfo \citep{Ahumada_2026}. When the viewing angle approaches the polar axis (i.e., $\theta_{\rm obs} < 60^\circ$), most combinations of $M_{\rm dyn}$ and $M_{\rm wind}$ are ruled out by our follow-up observations. For non-extreme edge-on orientations, limited by the model grid resolution, the constraints on the ejecta masses are approximately $M_{\rm dyn} \lesssim 0.01\,M_\odot$ and $M_{\rm wind} \lesssim 0.05\,M_\odot$. Allowing all model parameters to vary freely yields the results shown in Figure \ref{fig:All_constrain} of the Appendix. Although the constraining power is somewhat weakened in this broader parameter space, regions characterized by large ejecta masses, low velocities, high electron fractions, and small viewing angles remain disfavored by the observations.

    \subsection{Phenomenological Transient Model}
        To obtain more generalized constraints on the counterpart properties, we adopt a simplified profile to describe the photometric evolution of the counterpart. The absolute-magnitude light curve is defined as:
        \begin{equation}
            M(t)=M_{\rm peak,abs}
            +\Delta m_{1/2}
            \begin{cases}
            \displaystyle
            \left(
            \frac{t_{\rm peak}-t}{T_{\rm rise,1/2}}
            \right)^{\alpha_r},
            & t\leq t_{\rm peak},
            \\[12pt]
            \displaystyle
            \left(
            \frac{t-t_{\rm peak}}{T_{\rm decay,1/2}}
            \right)^{\alpha_d},
            & t>t_{\rm peak}.
            \end{cases}
            \label{eq:phenomenological_lc}
        \end{equation}
        The magnitude difference $\Delta m_{1/2}$ corresponding to half of the peak luminosity is $\sim0.753~{\rm mag}$. $T_{\rm rise,1/2}$ and $T_{\rm decay,1/2}$ correspond exactly to the time intervals between peak and half-peak luminosity on the rising and declining sides, respectively. We adopt $\alpha_r=2$ to describe a smooth, accelerating rise and $\alpha_d=1$ to represent an exponential luminosity decline. This simple asymmetric model broadly captures the typical evolution of many observed single-peaked optical transients. Because the adopted analytical function does not possess a finite physical explosion time, we introduce an operational explosion time. We set the explosion time to the epoch when the rising light curve is 5 mag fainter than the peak (corresponding to 1\% of the peak luminosity). For each injected counterpart, the operational explosion time is fixed to the GW merger time, which determines the corresponding peak epoch for a given rise timescale. Furthermore, the counterpart is assumed to have identical absolute magnitudes across the $g$, $r$, and $i$ bands. This simplification allows us to focus on how peak luminosity and evolution timescale affect detectability, although real transients generally have nonzero colors and color evolution.
        
\begin{figure}[htbp]
        \centering
        \begin{overpic}[width=0.45\textwidth]{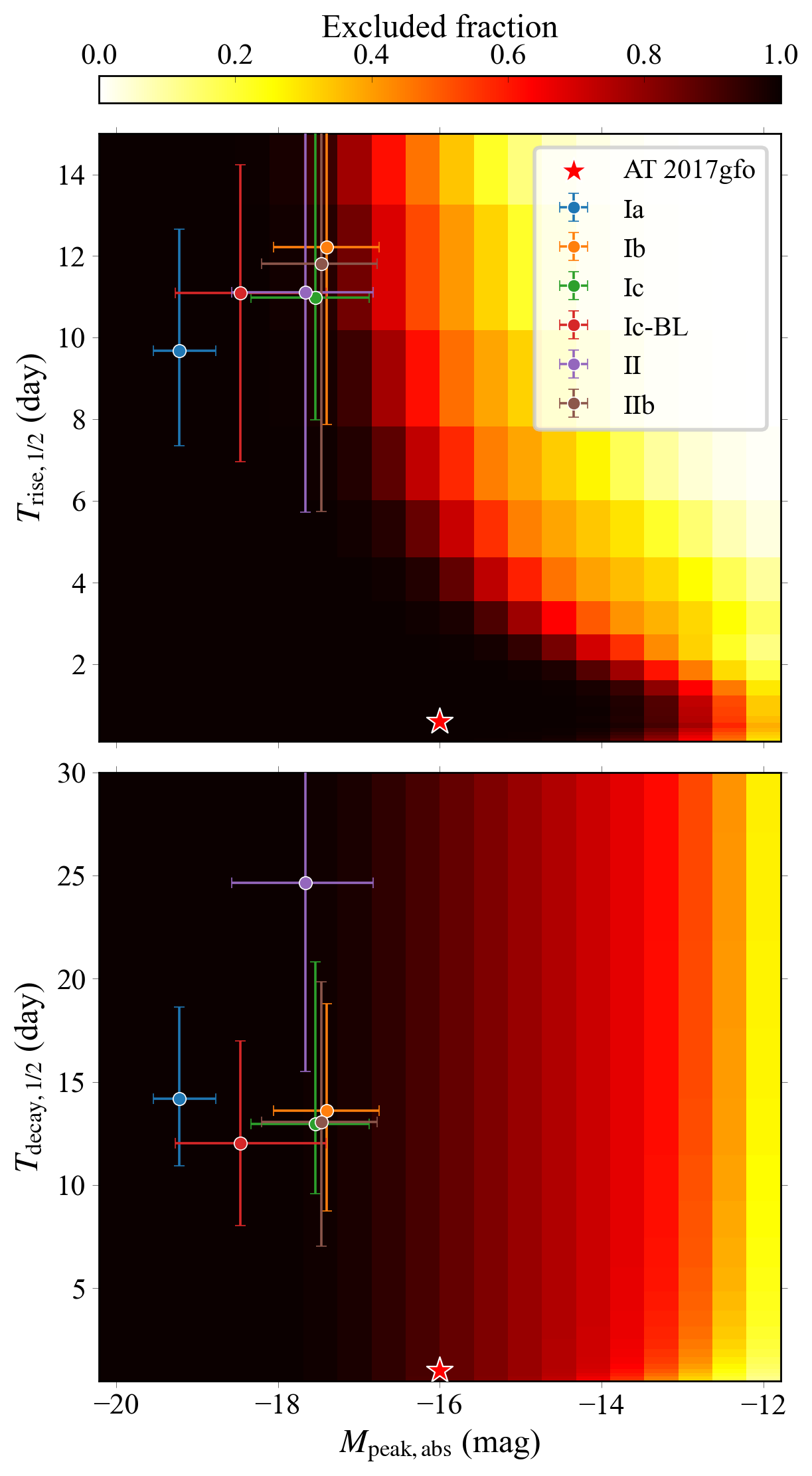}
        \end{overpic}
        \caption{Constraints on the peak absolute magnitude and the half-luminosity rising and decaying timescales. The color scale indicates the grid excluded fraction of the parameter space. Parameters of AT~2017gfo and several SN subtypes are also shown for comparison.}
        \label{fig:general_constraint}
        \end{figure}

        Constraints on the peak absolute magnitude and evolution timescale of the counterpart are shown in Figure \ref{fig:general_constraint}. We adopt the same exclusion criterion as in the BNS merger analysis, requiring at least two detections. The parameters of AT~2017gfo and several SN subtypes are also shown for comparison. The parameter estimates for different SN types are based on the ZTF Bright Transient Survey SN sample \citep{Fremling_2020,Perley_2020}. As shown in the upper panel of Figure \ref{fig:general_constraint}, the constraints on the peak absolute magnitude are stronger for rapidly rising transients due to the short, three-day duration of the WFST follow-up observations. At a given peak luminosity, the constraints become weaker for more slowly rising transients. In the region with $T_{\rm rise,1/2} > 8~\mathrm{days}$ and $M_{\rm peak,abs} > -15~\mathrm{mag}$, the constraints are dominated by the DECam observations due to their longer follow-up duration. Similarly, the model light curves remain in the rising phase for most parameter combinations due to the relatively short follow-up duration. Consequently, the decline timescale is only weakly constrained, as shown in the lower panel of Figure \ref{fig:general_constraint}. For rapidly evolving transients with $T_{\rm rise,1/2} \leq 2~\mathrm{days}$, the constraints on the peak absolute magnitude extend to $M_{\rm peak,abs} \approx -13~\mathrm{mag}$. For more slowly evolving transients, the observations also disfavor a substantial fraction of bright SN-like models that are assumed to begin at the GW merger time. These constraints do not apply to supernova emission that began before the GW trigger.
    
\section{EM Counterparts beyond the Present Search} \label{sec_6}

    The constraints presented in this work apply to rapidly evolving, off-nuclear optical transients emerging after the GW trigger. Other possible counterparts may occur in different environments or evolve on longer timescales and are therefore not constrained by our work.

    \subsection{Binary Black Hole Mergers in AGN Disks}
        BBH mergers embedded in AGN disks may produce EM emission through the interaction of the merger remnant with the surrounding gas. For example, \citet{Chen_2024c} considered emission powered by a kicked remnant accreting material from an AGN disk. Under fixed environmental conditions, the Bondi--Hoyle--Lyttleton accretion rate approximately scales as $\dot{M}_{\rm BHL}\propto M_{\rm rem}^{2}$. Taking a characteristic luminosity of $\sim10^{44}~{\rm erg~s^{-1}}$ for a $100\,M_{\odot}$ remnant as a reference, a simple extrapolation gives
        \begin{equation}
            L \sim 10^{40}
            \left(\frac{M_{\rm rem}}{1\,M_{\odot}}\right)^2
            {\rm erg~s^{-1}}
        \end{equation}
        for a solar-mass remnant. This estimate is only illustrative, since the radiative efficiency, jet formation, and surrounding disk properties may not preserve this simple mass scaling.
        
        Such emission would generally be several orders of magnitude fainter than the underlying AGN and could easily be hidden by intrinsic nuclear variability. Distinguishing it from ordinary AGN variability would therefore be difficult and would require dedicated long-term and multi-wavelength monitoring. 
    
    \subsection{Superkilonova-like Counterparts}
        Subsolar-mass neutron stars may form through fragmentation in the accretion disk produced during the collapse of a rapidly rotating massive star \citep{Metzger_2024}. Their subsequent merger could inject additional energy and $r$-process material into the supernova ejecta, producing a superkilonova-like event. However, the delay between core collapse and the compact-object merger depends on the formation and migration of the fragments and thus remains uncertain. It is also unclear whether this channel is preferentially associated with a particular core-collapse SN subtype.
        
        Observationally, the emission may be dominated by the expanding SN ejecta and show no unique signature that clearly distinguishes it from an ordinary CCSN. Although SN~2025ulz and SN~2025adtq were both classified as Type~IIb SNe and were found in temporal and spatial coincidence with subsolar-mass GW candidates, their physical associations remain inconclusive \citep{Kasliwal_2025,Hall_2026}. A systematic search for such events requires deep pre-merger imaging and longer-term monitoring, which are not available over most of our observed region. Future samples of nearby CCSNe with high-cadence pre-explosion observations, together with additional subsolar-mass GW candidates, may allow this scenario to be tested statistically.

\section{Conclusion} \label{sec_7}
    We have presented an optical follow-up search for an EM counterpart to the subsolar-mass compact-binary merger candidate S251112cm. Using wide coverage from WFST, together with the DECam observations presented by \citet{Hall_2026}, we searched for newly emerging, rapidly evolving, off-nuclear optical transients within the GW localization region. The combined WFST and DECam observations covered approximately $67\%$ of the localization probability in the updated skymap. No convincing EM counterpart was identified, and we use the observations to constrain both kilonova models and a phenomenological model of rapidly evolving optical transients.

    After applying our alert-selection procedure, we identified four transients whose host-galaxy distances are broadly consistent with the GW luminosity-distance posterior. However, all four sources are fainter and evolve substantially more slowly than AT~2017gfo, disfavoring their interpretation as rapidly evolving kilonova counterparts to S251112cm. Under the assumption that the counterpart was located within the observed footprint, the combined observations exclude a large fraction of the BNS kilonova model grid, particularly for viewing angles $\theta_{\rm obs}<60^{\circ}$. When the remaining parameters are fixed to the best-fitting values for AT~2017gfo, models with approximately $M_{\rm dyn}\gtrsim0.01\,M_{\odot}$ or $M_{\rm wind}\gtrsim0.05\,M_{\odot}$ are disfavored over most of the covered GW probability. Our phenomenological analysis also constrains rapidly rising transients with $T_{\rm rise,1/2}\leq2$ days to peak absolute magnitudes as faint as approximately $-13$ mag. These results are conditional on the adopted model grid, light-curve assumptions, and spatial coverage, and do not constrain nuclear AGN-disk emission or superkilonova-like events beginning before the GW trigger.

    Continued observations by the current GW detector network, followed by future improvements in detector sensitivity and localization, should produce a larger and better-characterized sample of subsolar-mass merger candidates. At the same time, the wide field, depth, and high survey cadence of WFST and Rubin-LSST will enable more complete searches for faint and rapidly evolving optical counterparts. Coordinated GW searches, rapid wide-field follow-up, deep pre-merger imaging, and long-term multi-wavelength monitoring will be essential for distinguishing genuine counterparts from unrelated transients and for determining the nature of subsolar-mass compact objects.

%
\facilities{WFST, DECam}


\section*{Acknowledgments}
The Wide Field Survey Telescope (WFST) is a joint facility of the University of Science and Technology of China, Purple Mountain Observatory. We appreciate the members of the WFST operation and maintenance team for their support. WZ is supported by the National Natural Science
Foundation of China (grant No. 12325301), Strategic Priority Research Program of the Chinese
Academy of Science (grant No. XDB0550300) and Fundamental and Interdisciplinary Disciplines Breakthrough Plan of Ministry of Education of China. ZYL acknowledges support from the Postdoctoral Fellowship Program of the China Postdoctoral Science Foundation (grant No. GZC20261722).

\software{Astropy \citep{astropy:2013, astropy:2018, astropy:2022}; SciPy \citep{2020SciPy-NMeth}; NumPy \citep{harris2020array}; Matplotlib \citep{Hunter:2007}; SWarp \citep{SWarp}; dustmaps \citep{Green2018}; ligo.skymap \citep{Singer_2016_a, Singer_2016_b}; HOTPANTS \citep{HOTPANTS}; dynesty \citep{2004AIPC..735..395S,2020MNRAS.493.3132S,sergey_koposov_2025_17268284}}


\appendix

\section{Constraints on the Full BNS Model Grid}
\begin{figure*}[htbp]
    \centering
    \begin{overpic}[width=0.8\textwidth]{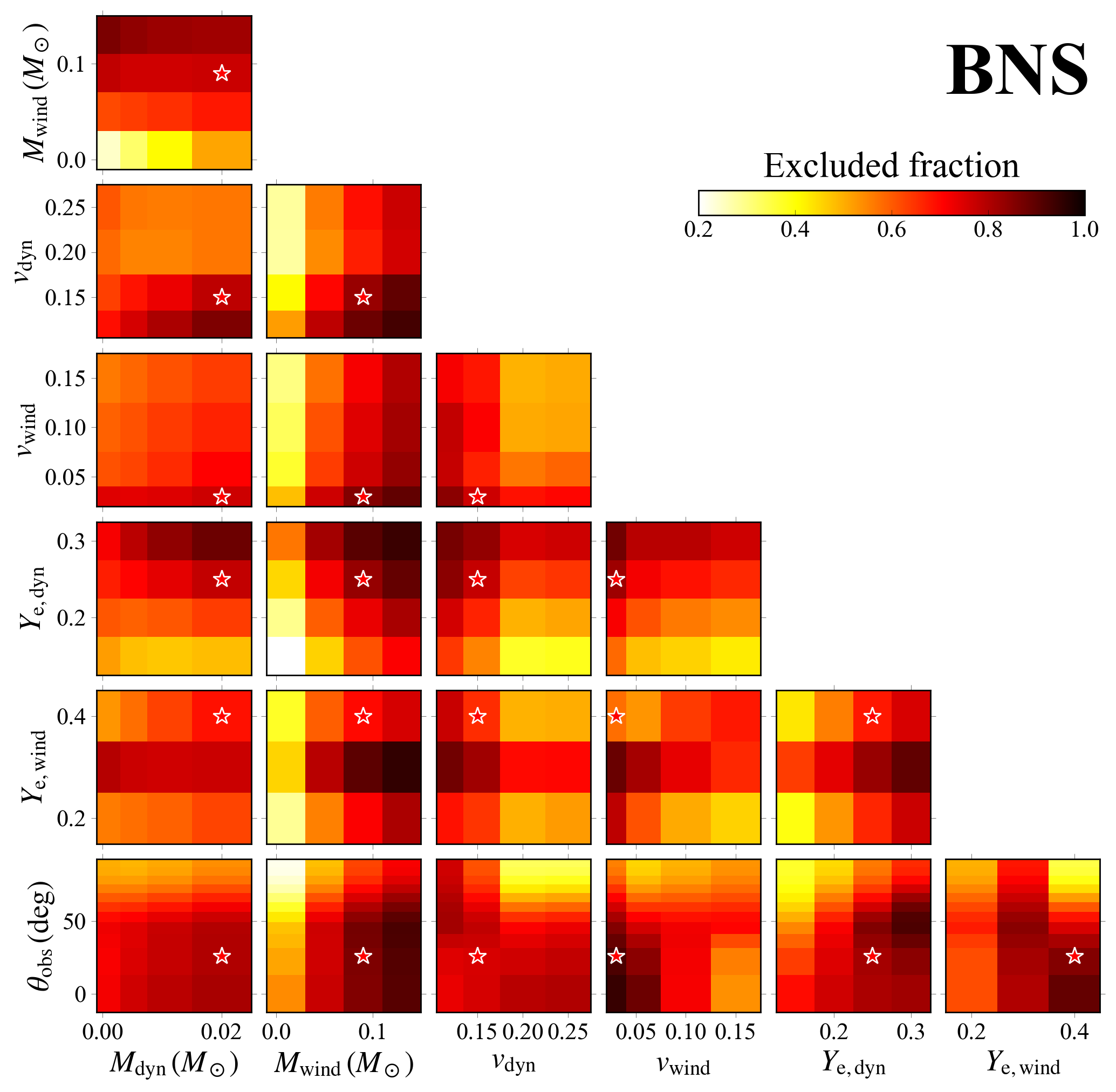}
    \end{overpic}
    \caption{Constraints on the full parameter space of the POSSIS BNS model grid presented by \citet{Ahumada_2026}. The color scale indicates the fraction of the model grid excluded for each parameter combination. Red stars mark the inferred ejecta masses and viewing angle of AT~2017gfo for comparison.}
    \label{fig:All_constrain}
\end{figure*}

This appendix presents constraints on the full parameter space of the POSSIS BNS model grid from \citet{Ahumada_2026}. Figure~\ref{fig:All_constrain} shows the excluded fraction of the model grid for different parameter combinations.


\bibliography{main}{}

@ARTICLE{2025GCN.42690....1L,
       author = {{Ligo Scientific Collaboration} and {VIRGO Collaboration} and {Kagra Collaboration}},
        title = "{LIGO/Virgo/KAGRA S251112cm: Updated Sky localization and false alarm rate estimate}",
      journal = {GRB Coordinates Network},
         year = 2025,
        month = nov,
       volume = {42690},
        pages = {1},
       adsurl = {https://ui.adsabs.harvard.edu/abs/2025GCN.42690....1L}
}

@article{Schlafly_2011,
  title = {{{MEASURING REDDENING WITH SLOAN DIGITAL SKY SURVEY STELLAR SPECTRA AND RECALIBRATING SFD}}},
  author = {Schlafly, Edward F. and Finkbeiner, Douglas P.},
  year = {2011},
  month = aug,
  journal = {The Astrophysical Journal},
  volume = {737},
  number = {2},
  pages = {103},
  issn = {0004-637X, 1538-4357},
  doi = {10.1088/0004-637X/737/2/103},
  urldate = {2025-05-10}
}

@article{Liu_2026a,
  title = {Illuminating the {{Mass Gap}} through a {{Deep Optical Constraint}} on the {{Neutron Star Merger Candidate S250206dm}}},
  author = {Liu, Zhengyan and Xu, Zelin and Jiang, Ji-an and Zhao, Wen and Jin, Zhiping and Dai, Zigao and Meng, Dezheng and Wu, Xuefeng and Wei, Daming and Liang, Runduo and He, Lei and Cai, Minxuan and Fan, Lulu and Wu, Weiyu and Zhao, Junhan and Jia, Ziqing and Yu, Kexin and Geng, Jinjun and Xiao, Di and Li, Feng and Tang, Jinlong and Zuo, Yingxi and Zhang, Xiaoling and Liu, Hao and Wang, Jian and Zhang, Hongfei and Liang, Ming and Wang, Hairen and Yao, Dazhi and Hu, Lei and Kong, Xu and Li, Bin and Jiang, Ning and Wang, Tinggui and Wan, Zhen and Xue, Yongquan and Zhu, Qingfeng and Zheng, Xianzhong},
  year = 2026,
  month = mar,
  journal = {The Astrophysical Journal Letters},
  volume = {1000},
  number = {1},
  pages = {L20},
  issn = {2041-8205, 2041-8213},
  doi = {10.3847/2041-8213/ae4d14},
  urldate = {2026-03-16}
}

@article{Deng_2021,
  title = {Lenghu on the {{Tibetan Plateau}} as an Astronomical Observing Site},
  author = {Deng, Licai and Yang, Fan and Chen, Xiaodian and He, Fei and Liu, Qili and Zhang, Bo and Zhang, Chunguang and Wang, Kun and Liu, Nian and Ren, Anbing and Luo, Zhiquan and Yan, Zhengzhou and Tian, Jianfeng and Pan, Jun},
  year = {2021},
  month = aug,
  journal = {Nature},
  volume = {596},
  number = {7872},
  pages = {353--356},
  issn = {0028-0836, 1476-4687},
  doi = {10.1038/s41586-021-03711-z},
  urldate = {2023-10-30}
}

@article{Wang_2023,
  title = {Science with the 2.5-Meter {{Wide Field Survey Telescope}} ({{WFST}})},
  author = {Wang, Tinggui and Liu, Guilin and Cai, Zhenyi and Geng, Jinjun and Fang, Min and He, Haoning and Jiang, Ji-an and Jiang, Ning and Kong, Xu and Li, Bin and Li, Ye and Luo, Wentao and Pan, Zhizheng and Wu, Xuefeng and Yang, Ji and Yu, Jiming and Zheng, Xianzhong and Zhu, Qingfeng and Cai, Yi-Fu and Chen, Yuanyuan and Chen, Zhiwei and Dai, Zigao and Fan, Lulu and Fan, Yizhong and Fang, Wenjuan and He, Zhicheng and Hu, Lei and Hu, Maokai and Jin, Zhiping and Jiang, Zhibo and Li, Guoliang and Li, Fan and Li, Xuzhi and Liang, Runduo and Lin, Zheyu and Liu, Qingzhong and Liu, Wenhao and Liu, Zhengyan and Liu, Wei and Liu, Yao and Lou, Zheng and Qu, Han and Sheng, Zhenfeng and Shi, Jianchun and Shu, Yiping and Su, Zhenbo and Sun, Tianrui and Wang, Hongchi and Wang, Huiyuan and Wang, Jian and Wang, Junxian and Wei, Daming and Wei, Junjie and Xue, Yongquan and Yan, Jingzhi and Yang, Chao and Yuan, Ye and Yuan, Yefei and Zhang, Hongxin and Zhang, Miaomiao and Zhao, Haibin and Zhao, Wen},
  year = {2023},
  month = oct,
  journal = {Science China Physics, Mechanics \& Astronomy},
  volume = {66},
  number = {10},
  pages = {109512},
  issn = {1674-7348, 1869-1927},
  doi = {10.1007/s11433-023-2197-5},
  urldate = {2023-10-18}
}

@article{conroy2023china,
  title={China's powerful new telescope will search for exploding stars.},
  author={Conroy, Gemma},
  journal={Nature},
  year={2023},
  month = sep,
  doi = {10.1038/d41586-023-03013-6}
}

@article{Liu_2023,
  title = {Target-of-{{Opportunity Observation Detectability}} of {{Kilonovae}} with {{WFST}}},
  author = {Liu, Zheng-Yan and Lin, Zhe-Yu and Yu, Ji-Ming and Wang, Hui-Yu and Mourani, Gibran-Marc and Zhao, Wen and Dai, Zi-Gao},
  year = {2023},
  month = apr,
  journal = {The Astrophysical Journal},
  volume = {947},
  number = {2},
  pages = {59},
  issn = {0004-637X, 1538-4357},
  doi = {10.3847/1538-4357/acc73b},
  urldate = {2023-05-23}
}

@ARTICLE{2025GCN.42650....1L,
       author = {{Ligo Scientific Collaboration} and {VIRGO Collaboration} and {Kagra Collaboration}},
        title = "{LIGO/Virgo/KAGRA S251112cm: Identification of a GW compact binary merger candidate}",
      journal = {GRB Coordinates Network},
         year = 2025,
        month = nov,
       volume = {42650},
        pages = {1},
       adsurl = {https://ui.adsabs.harvard.edu/abs/2025GCN.42650....1L}
}

@article{astropy:2013,
Adsurl = {http://adsabs.harvard.edu/abs/2013A%26A...558A..33A},
Archiveprefix = {arXiv},
Author = {{Astropy Collaboration} and {Robitaille}, T.~P. and {Tollerud}, E.~J. and {Greenfield}, P. and {Droettboom}, M. and {Bray}, E. and {Aldcroft}, T. and {Davis}, M. and {Ginsburg}, A. and {Price-Whelan}, A.~M. and {Kerzendorf}, W.~E. and {Conley}, A. and {Crighton}, N. and {Barbary}, K. and {Muna}, D. and {Ferguson}, H. and {Grollier}, F. and {Parikh}, M.~M. and {Nair}, P.~H. and {Unther}, H.~M. and {Deil}, C. and {Woillez}, J. and {Conseil}, S. and {Kramer}, R. and {Turner}, J.~E.~H. and {Singer}, L. and {Fox}, R. and {Weaver}, B.~A. and {Zabalza}, V. and {Edwards}, Z.~I. and {Azalee Bostroem}, K. and {Burke}, D.~J. and {Casey}, A.~R. and {Crawford}, S.~M. and {Dencheva}, N. and {Ely}, J. and {Jenness}, T. and {Labrie}, K. and {Lim}, P.~L. and {Pierfederici}, F. and {Pontzen}, A. and {Ptak}, A. and {Refsdal}, B. and {Servillat}, M. and {Streicher}, O.},
Doi = {10.1051/0004-6361/201322068},
Eid = {A33},
Eprint = {1307.6212},
Journal = {\aap},
Month = oct,
Pages = {A33},
Primaryclass = {astro-ph.IM},
Title = {{Astropy: A community Python package for astronomy}},
Volume = 558,
Year = 2013}

@ARTICLE{astropy:2018,
       author = {{Astropy Collaboration} and {Price-Whelan}, A.~M. and
         {Sip{\H{o}}cz}, B.~M. and {G{\"u}nther}, H.~M. and {Lim}, P.~L. and
         {Crawford}, S.~M. and {Conseil}, S. and {Shupe}, D.~L. and
         {Craig}, M.~W. and {Dencheva}, N. and {Ginsburg}, A. and {Vand
        erPlas}, J.~T. and {Bradley}, L.~D. and {P{\'e}rez-Su{\'a}rez}, D. and
         {de Val-Borro}, M. and {Aldcroft}, T.~L. and {Cruz}, K.~L. and
         {Robitaille}, T.~P. and {Tollerud}, E.~J. and {Ardelean}, C. and
         {Babej}, T. and {Bach}, Y.~P. and {Bachetti}, M. and {Bakanov}, A.~V. and
         {Bamford}, S.~P. and {Barentsen}, G. and {Barmby}, P. and
         {Baumbach}, A. and {Berry}, K.~L. and {Biscani}, F. and {Boquien}, M. and
         {Bostroem}, K.~A. and {Bouma}, L.~G. and {Brammer}, G.~B. and
         {Bray}, E.~M. and {Breytenbach}, H. and {Buddelmeijer}, H. and
         {Burke}, D.~J. and {Calderone}, G. and {Cano Rodr{\'\i}guez}, J.~L. and
         {Cara}, M. and {Cardoso}, J.~V.~M. and {Cheedella}, S. and {Copin}, Y. and
         {Corrales}, L. and {Crichton}, D. and {D'Avella}, D. and {Deil}, C. and
         {Depagne}, {\'E}. and {Dietrich}, J.~P. and {Donath}, A. and
         {Droettboom}, M. and {Earl}, N. and {Erben}, T. and {Fabbro}, S. and
         {Ferreira}, L.~A. and {Finethy}, T. and {Fox}, R.~T. and
         {Garrison}, L.~H. and {Gibbons}, S.~L.~J. and {Goldstein}, D.~A. and
         {Gommers}, R. and {Greco}, J.~P. and {Greenfield}, P. and
         {Groener}, A.~M. and {Grollier}, F. and {Hagen}, A. and {Hirst}, P. and
         {Homeier}, D. and {Horton}, A.~J. and {Hosseinzadeh}, G. and {Hu}, L. and
         {Hunkeler}, J.~S. and {Ivezi{\'c}}, {\v{Z}}. and {Jain}, A. and
         {Jenness}, T. and {Kanarek}, G. and {Kendrew}, S. and {Kern}, N.~S. and
         {Kerzendorf}, W.~E. and {Khvalko}, A. and {King}, J. and {Kirkby}, D. and
         {Kulkarni}, A.~M. and {Kumar}, A. and {Lee}, A. and {Lenz}, D. and
         {Littlefair}, S.~P. and {Ma}, Z. and {Macleod}, D.~M. and
         {Mastropietro}, M. and {McCully}, C. and {Montagnac}, S. and
         {Morris}, B.~M. and {Mueller}, M. and {Mumford}, S.~J. and {Muna}, D. and
         {Murphy}, N.~A. and {Nelson}, S. and {Nguyen}, G.~H. and
         {Ninan}, J.~P. and {N{\"o}the}, M. and {Ogaz}, S. and {Oh}, S. and
         {Parejko}, J.~K. and {Parley}, N. and {Pascual}, S. and {Patil}, R. and
         {Patil}, A.~A. and {Plunkett}, A.~L. and {Prochaska}, J.~X. and
         {Rastogi}, T. and {Reddy Janga}, V. and {Sabater}, J. and
         {Sakurikar}, P. and {Seifert}, M. and {Sherbert}, L.~E. and
         {Sherwood-Taylor}, H. and {Shih}, A.~Y. and {Sick}, J. and
         {Silbiger}, M.~T. and {Singanamalla}, S. and {Singer}, L.~P. and
         {Sladen}, P.~H. and {Sooley}, K.~A. and {Sornarajah}, S. and
         {Streicher}, O. and {Teuben}, P. and {Thomas}, S.~W. and
         {Tremblay}, G.~R. and {Turner}, J.~E.~H. and {Terr{\'o}n}, V. and
         {van Kerkwijk}, M.~H. and {de la Vega}, A. and {Watkins}, L.~L. and
         {Weaver}, B.~A. and {Whitmore}, J.~B. and {Woillez}, J. and
         {Zabalza}, V. and {Astropy Contributors}},
        title = "{The Astropy Project: Building an Open-science Project and Status of the v2.0 Core Package}",
      journal = {\aj},
         year = 2018,
        month = sep,
       volume = {156},
       number = {3},
          eid = {123},
        pages = {123},
          doi = {10.3847/1538-3881/aabc4f},
archivePrefix = {arXiv},
       eprint = {1801.02634},
 primaryClass = {astro-ph.IM},
       adsurl = {https://ui.adsabs.harvard.edu/abs/2018AJ....156..123A}
}

@ARTICLE{astropy:2022,
       author = {{Astropy Collaboration} and {Price-Whelan}, Adrian M. and {Lim}, Pey Lian and {Earl}, Nicholas and {Starkman}, Nathaniel and {Bradley}, Larry and {Shupe}, David L. and {Patil}, Aarya A. and {Corrales}, Lia and {Brasseur}, C.~E. and {N{"o}the}, Maximilian and {Donath}, Axel and {Tollerud}, Erik and {Morris}, Brett M. and {Ginsburg}, Adam and {Vaher}, Eero and {Weaver}, Benjamin A. and {Tocknell}, James and {Jamieson}, William and {van Kerkwijk}, Marten H. and {Robitaille}, Thomas P. and {Merry}, Bruce and {Bachetti}, Matteo and {G{"u}nther}, H. Moritz and {Aldcroft}, Thomas L. and {Alvarado-Montes}, Jaime A. and {Archibald}, Anne M. and {B{'o}di}, Attila and {Bapat}, Shreyas and {Barentsen}, Geert and {Baz{'a}n}, Juanjo and {Biswas}, Manish and {Boquien}, M{'e}d{'e}ric and {Burke}, D.~J. and {Cara}, Daria and {Cara}, Mihai and {Conroy}, Kyle E. and {Conseil}, Simon and {Craig}, Matthew W. and {Cross}, Robert M. and {Cruz}, Kelle L. and {D'Eugenio}, Francesco and {Dencheva}, Nadia and {Devillepoix}, Hadrien A.~R. and {Dietrich}, J{"o}rg P. and {Eigenbrot}, Arthur Davis and {Erben}, Thomas and {Ferreira}, Leonardo and {Foreman-Mackey}, Daniel and {Fox}, Ryan and {Freij}, Nabil and {Garg}, Suyog and {Geda}, Robel and {Glattly}, Lauren and {Gondhalekar}, Yash and {Gordon}, Karl D. and {Grant}, David and {Greenfield}, Perry and {Groener}, Austen M. and {Guest}, Steve and {Gurovich}, Sebastian and {Handberg}, Rasmus and {Hart}, Akeem and {Hatfield-Dodds}, Zac and {Homeier}, Derek and {Hosseinzadeh}, Griffin and {Jenness}, Tim and {Jones}, Craig K. and {Joseph}, Prajwel and {Kalmbach}, J. Bryce and {Karamehmetoglu}, Emir and {Ka{l}uszy{'n}ski}, Miko{l}aj and {Kelley}, Michael S.~P. and {Kern}, Nicholas and {Kerzendorf}, Wolfgang E. and {Koch}, Eric W. and {Kulumani}, Shankar and {Lee}, Antony and {Ly}, Chun and {Ma}, Zhiyuan and {MacBride}, Conor and {Maljaars}, Jakob M. and {Muna}, Demitri and {Murphy}, N.~A. and {Norman}, Henrik and {O'Steen}, Richard and {Oman}, Kyle A. and {Pacifici}, Camilla and {Pascual}, Sergio and {Pascual-Granado}, J. and {Patil}, Rohit R. and {Perren}, Gabriel I. and {Pickering}, Timothy E. and {Rastogi}, Tanuj and {Roulston}, Benjamin R. and {Ryan}, Daniel F. and {Rykoff}, Eli S. and {Sabater}, Jose and {Sakurikar}, Parikshit and {Salgado}, Jes{'u}s and {Sanghi}, Aniket and {Saunders}, Nicholas and {Savchenko}, Volodymyr and {Schwardt}, Ludwig and {Seifert-Eckert}, Michael and {Shih}, Albert Y. and {Jain}, Anany Shrey and {Shukla}, Gyanendra and {Sick}, Jonathan and {Simpson}, Chris and {Singanamalla}, Sudheesh and {Singer}, Leo P. and {Singhal}, Jaladh and {Sinha}, Manodeep and {Sip{H{o}}cz}, Brigitta M. and {Spitler}, Lee R. and {Stansby}, David and {Streicher}, Ole and {{{S}}umak}, Jani and {Swinbank}, John D. and {Taranu}, Dan S. and {Tewary}, Nikita and {Tremblay}, Grant R. and {Val-Borro}, Miguel de and {Van Kooten}, Samuel J. and {Vasovi{'c}}, Zlatan and {Verma}, Shresth and {de Miranda Cardoso}, Jos{'e} Vin{'i}cius and {Williams}, Peter K.~G. and {Wilson}, Tom J. and {Winkel}, Benjamin and {Wood-Vasey}, W.~M. and {Xue}, Rui and {Yoachim}, Peter and {Zhang}, Chen and {Zonca}, Andrea and {Astropy Project Contributors}},
        title = "{The Astropy Project: Sustaining and Growing a Community-oriented Open-source Project and the Latest Major Release (v5.0) of the Core Package}",
      journal = {\apj},
         year = 2022,
        month = aug,
       volume = {935},
       number = {2},
          eid = {167},
        pages = {167},
          doi = {10.3847/1538-4357/ac7c74},
archivePrefix = {arXiv},
       eprint = {2206.14220},
 primaryClass = {astro-ph.IM},
       adsurl = {https://ui.adsabs.harvard.edu/abs/2022ApJ...935..167A}
}

@software{HOTPANTS,
       author = {{Becker}, Andrew},
        title = "{HOTPANTS: High Order Transform of PSF ANd Template Subtraction}",
 howpublished = {Astrophysics Source Code Library, record ascl:1504.004},
         year = 2015,
        month = apr,
          eid = {ascl:1504.004},
archivePrefix = {ascl},
       eprint = {1504.004},
       adsurl = {https://ui.adsabs.harvard.edu/abs/2015ascl.soft04004B}
}

@ARTICLE{2020SciPy-NMeth,
  author  = {Virtanen, Pauli and Gommers, Ralf and Oliphant, Travis E. and
            Haberland, Matt and Reddy, Tyler and Cournapeau, David and
            Burovski, Evgeni and Peterson, Pearu and Weckesser, Warren and
            Bright, Jonathan and {van der Walt}, St{\'e}fan J. and
            Brett, Matthew and Wilson, Joshua and Millman, K. Jarrod and
            Mayorov, Nikolay and Nelson, Andrew R. J. and Jones, Eric and
            Kern, Robert and Larson, Eric and Carey, C J and
            Polat, {\.I}lhan and Feng, Yu and Moore, Eric W. and
            {VanderPlas}, Jake and Laxalde, Denis and Perktold, Josef and
            Cimrman, Robert and Henriksen, Ian and Quintero, E. A. and
            Harris, Charles R. and Archibald, Anne M. and
            Ribeiro, Ant{\^o}nio H. and Pedregosa, Fabian and
            {van Mulbregt}, Paul and {SciPy 1.0 Contributors}},
  title   = {{{SciPy} 1.0: Fundamental Algorithms for Scientific
            Computing in Python}},
  journal = {Nature Methods},
  year    = {2020},
  volume  = {17},
  pages   = {261--272},
  adsurl  = {https://rdcu.be/b08Wh},
  doi     = {10.1038/s41592-019-0686-2},
}

@Article{harris2020array,
 title         = {Array programming with {NumPy}},
 author        = {Charles R. Harris and K. Jarrod Millman and St{\'{e}}fan J.
                 van der Walt and Ralf Gommers and Pauli Virtanen and David
                 Cournapeau and Eric Wieser and Julian Taylor and Sebastian
                 Berg and Nathaniel J. Smith and Robert Kern and Matti Picus
                 and Stephan Hoyer and Marten H. van Kerkwijk and Matthew
                 Brett and Allan Haldane and Jaime Fern{\'{a}}ndez del
                 R{\'{i}}o and Mark Wiebe and Pearu Peterson and Pierre
                 G{\'{e}}rard-Marchant and Kevin Sheppard and Tyler Reddy and
                 Warren Weckesser and Hameer Abbasi and Christoph Gohlke and
                 Travis E. Oliphant},
 year          = {2020},
 month         = sep,
 journal       = {Nature},
 volume        = {585},
 number        = {7825},
 pages         = {357--362},
 doi           = {10.1038/s41586-020-2649-2},
 publisher     = {Springer Science and Business Media {LLC}},
 url           = {https://doi.org/10.1038/s41586-020-2649-2}
}

@Article{Hunter:2007,
  Author    = {Hunter, J. D.},
  Title     = {Matplotlib: A 2D graphics environment},
  Journal   = {Computing in Science \& Engineering},
  Volume    = {9},
  Number    = {3},
  Pages     = {90--95},
  publisher = {IEEE COMPUTER SOC},
  doi       = {10.1109/MCSE.2007.55},
  year      = 2007
}

@software{SWarp,
       author = {{Bertin}, Emmanuel},
        title = "{SWarp: Resampling and Co-adding FITS Images Together}",
 howpublished = {Astrophysics Source Code Library, record ascl:1010.068},
         year = 2010,
        month = oct,
          eid = {ascl:1010.068},
archivePrefix = {ascl},
       eprint = {1010.068},
       adsurl = {https://ui.adsabs.harvard.edu/abs/2010ascl.soft10068B}
}

@article{Green2018, doi = {10.21105/joss.00695}, url = {https://doi.org/10.21105/joss.00695}, year = {2018}, publisher = {The Open Journal}, volume = {3}, number = {26}, pages = {695}, author = {Green, Gregory M.}, title = {dustmaps: A Python interface for maps of interstellar dust}, journal = {Journal of Open Source Software} }

@INPROCEEDINGS{2004AIPC..735..395S,
       author = {{Skilling}, John},
        title = "{Nested Sampling}",
    booktitle = {Bayesian Inference and Maximum Entropy Methods in Science and Engineering: 24th International Workshop on Bayesian Inference and Maximum Entropy Methods in Science and Engineering},
         year = 2004,
       editor = {{Fischer}, Rainer and {Preuss}, Roland and {Toussaint}, Udo Von},
       series = {American Institute of Physics Conference Series},
       volume = {735},
        month = nov,
    publisher = {AIP},
        pages = {395-405},
          doi = {10.1063/1.1835238},
       adsurl = {https://ui.adsabs.harvard.edu/abs/2004AIPC..735..395S}
}

@ARTICLE{2020MNRAS.493.3132S,
       author = {{Speagle}, Joshua S.},
        title = "{DYNESTY: a dynamic nested sampling package for estimating Bayesian posteriors and evidences}",
      journal = {\mnras},
         year = 2020,
        month = apr,
       volume = {493},
       number = {3},
        pages = {3132-3158},
          doi = {10.1093/mnras/staa278},
archivePrefix = {arXiv},
       eprint = {1904.02180},
 primaryClass = {astro-ph.IM},
       adsurl = {https://ui.adsabs.harvard.edu/abs/2020MNRAS.493.3132S}
}

@software{sergey_koposov_2025_17268284,
  author       = {Sergey Koposov and
                  Josh Speagle and
                  Kyle Barbary and
                  Gregory Ashton and
                  Ed Bennett and
                  Johannes Buchner and
                  Carl Scheffler and
                  Colm Talbot and
                  Ben Cook and
                  James Guillochon and
                  Patricio Cubillos and
                  Andrés Asensio Ramos and
                  Matthieu Dartiailh and
                  Ilya and
                  Erik Tollerud and
                  Dustin Lang and
                  Ben Johnson and
                  jtmendel and
                  Edward Higson and
                  Thomas Vandal and
                  Tansu Daylan and
                  Ruth Angus and
                  patelR and
                  Phillip Cargile and
                  Patrick Sheehan and
                  Matt Pitkin and
                  Matthew Kirk and
                  Lu Xu and
                  Joel Leja and
                  joezuntz},
  title        = {joshspeagle/dynesty: v3.0.0},
  month        = oct,
  year         = 2025,
  publisher    = {Zenodo},
  version      = {v3.0.0},
  doi          = {10.5281/zenodo.17268284},
  url          = {https://doi.org/10.5281/zenodo.17268284},
  swhid        = {swh:1:dir:b3340f6aaa931bae6c6bcff6e7ddeb89ca508a15
                   ;origin=https://doi.org/10.5281/zenodo.3348367;vis
                   it=swh:1:snp:fa92422c27d5611b107216e9b766a871bb43b
                   bee;anchor=swh:1:rel:94af3a8ef6e50d997258e152c3833
                   3550bd7463f;path=joshspeagle-dynesty-217bc94
                  },
}

@article{Singer_2016_a,
doi = {10.3847/2041-8205/829/1/L15},
url = {https://doi.org/10.3847/2041-8205/829/1/L15},
year = {2016},
month = {sep},
publisher = {The American Astronomical Society},
volume = {829},
number = {1},
pages = {L15},
author = {Singer, Leo P. and Chen, Hsin-Yu and Holz, Daniel E. and Farr, Will M. and Price, Larry R. and Raymond, Vivien and Cenko, S. Bradley and Gehrels, Neil and Cannizzo, John and Kasliwal, Mansi M. and Nissanke, Samaya and Coughlin, Michael and Farr, Ben and Urban, Alex L. and Vitale, Salvatore and Veitch, John and Graff, Philip and Berry, Christopher P. L. and Mohapatra, Satya and Mandel, Ilya},
title = {GOING THE DISTANCE: MAPPING HOST GALAXIES OF LIGO AND VIRGO SOURCES IN THREE DIMENSIONS USING LOCAL COSMOGRAPHY AND TARGETED FOLLOW-UP},
journal = {The Astrophysical Journal Letters},
}

@article{Singer_2016_b,
doi = {10.3847/0067-0049/226/1/10},
url = {https://doi.org/10.3847/0067-0049/226/1/10},
year = {2016},
month = {sep},
publisher = {The American Astronomical Society},
volume = {226},
number = {1},
pages = {10},
author = {Singer, Leo P. and Chen, Hsin-Yu and Holz, Daniel E. and Farr, Will M. and Price, Larry R. and Raymond, Vivien and Cenko, S. Bradley and Gehrels, Neil and Cannizzo, John and Kasliwal, Mansi M. and Nissanke, Samaya and Coughlin, Michael and Farr, Ben and Urban, Alex L. and Vitale, Salvatore and Veitch, John and Graff, Philip and Berry, Christopher P. L. and Mohapatra, Satya and Mandel, Ilya},
title = {SUPPLEMENT: “GOING THE DISTANCE: MAPPING HOST GALAXIES OF LIGO AND VIRGO SOURCES IN THREE DIMENSIONS USING LOCAL COSMOGRAPHY AND TARGETED FOLLOW-UP” (2016, ApJL, 829, L15)},
journal = {The Astrophysical Journal Supplement Series},
}

@article{boschOverviewLSSTImage2018,
  title = {An {{Overview}} of the {{LSST Image Processing Pipelines}}},
  author = {Bosch, James and AlSayyad, Yusra and Armstrong, Robert and Bellm, Eric and Chiang, Hsin-Fang and Eggl, Siegfried and Findeisen, Krzysztof and {Fisher-Levine}, Merlin and Guy, Leanne P. and Guyonnet, Augustin and Ivezi{\'c}, {\v Z}eljko and Jenness, Tim and Kov{\'a}cs, G{\'a}bor and Krughoff, K. Simon and Lupton, Robert H. and Lust, Nate B. and MacArthur, Lauren A. and Meyers, Joshua and Moolekamp, Fred and Morrison, Christopher B. and Morton, Timothy D. and O'Mullane, William and Parejko, John K. and Plazas, Andr{\'e}s A. and Price, Paul A. and Rawls, Meredith L. and Reed, Sophie L. and Schellart, Pim and Slater, Colin T. and Sullivan, Ian and Swinbank, John D. and Taranu, Dan and Waters, Christopher Z. and {Wood-Vasey}, W. M.},
  year = {2018},
  month = dec,
  journal = {arXiv e-prints},
  number = {arXiv:1812.03248},
  eprint = {1812.03248},
  primaryclass = {astro-ph},
  publisher = {arXiv},
  urldate = {2024-06-26},
  archiveprefix = {arxiv},
  langid = {english},
}

@article{Hu_2022,
doi = {10.3847/1538-4357/ac7394},
url = {https://dx.doi.org/10.3847/1538-4357/ac7394},
year = {2022},
month = {sep},
publisher = {The American Astronomical Society},
volume = {936},
number = {2},
pages = {157},
author = {Hu, Lei and Wang, Lifan and Chen, Xingzhuo and Yang, Jiawen},
title = {Image Subtraction in Fourier Space},
journal = {The Astrophysical Journal},
}

@ARTICLE{Cai_2025,
  title = {The 2.5-Meter {{Wide Field Survey Telescope Real-time Data Processing Pipeline I}}: {{From}} Raw Data to Alert Distribution},
  shorttitle = {The 2.5-Meter {{Wide Field Survey Telescope Real-time Data Processing Pipeline I}}},
  author = {Cai, Minxuan and Xu, Zelin and Fan, Lulu and Wan, Zhen and Kong, Xu and Hu, Weida and Jiang, Ji-an and Hu, Lei and Zhu, Qing-feng and Li, Guoliang and Lin, Jie and Fang, Min and Xue, Yongquan and Zhen, Xianzhong and Wang, Tinggui},
  year = {2025},
  month = jan,
  journal = {arXiv e-prints},
  number = {arXiv:2501.15018},
  eprint = {2501.15018},
  primaryclass = {astro-ph},
  publisher = {arXiv},
  doi = {10.48550/arXiv.2501.15018},
  urldate = {2025-06-26},
  archiveprefix = {arXiv}
}

@ARTICLE{2025GCN.42722....1L,
       author = {{Liu}, Z.~Y. and {Zhao}, W. and {Jiang}, J.-A. and {Xu}, Z.~L. and {Meng}, D.~Z. and {Cai}, M.~X. and {Wang}, T.~G. and {Kong}, X. and {Dai}, Z.~G. and {Fan}, L.~L. and {Jin}, Z.~P. and {Fan}, Y.~Z. and {WFST Collaboration}},
        title = "{LIGO/Virgo/KAGRA S251112cm: WFST follow-up observations and preliminary results}",
      journal = {GRB Coordinates Network},
         year = 2025,
        month = nov,
       volume = {42722},
        pages = {1},
       adsurl = {https://ui.adsabs.harvard.edu/abs/2025GCN.42722....1L}
}

@article{Liu_2025,
  title = {Classification of Real and Bogus Transients Using Active Learning and Semi-Supervised Learning},
  author = {Liu, Yating and Fan, Lulu and Hu, Lei and Lu, Junqiang and Lu, Yan and Xu, Zelin and Zhu, Jiazheng and Wang, Haochen and Kong, Xu},
  year = {2025},
  month = jan,
  journal = {Astronomy \& Astrophysics},
  volume = {693},
  pages = {A105},
  publisher = {EDP Sciences},
  issn = {0004-6361, 1432-0746},
  doi = {10.1051/0004-6361/202348581},
  urldate = {2025-07-18},
  copyright = {https://creativecommons.org/licenses/by/4.0}
}

@ARTICLE{Chambers_2016,
       author = {{Chambers}, K.~C. and {Magnier}, E.~A. and {Metcalfe}, N. and {Flewelling}, H.~A. and {Huber}, M.~E. and {Waters}, C.~Z. and {Denneau}, L. and {Draper}, P.~W. and {Farrow}, D. and {Finkbeiner}, D.~P. and {Holmberg}, C. and {Koppenhoefer}, J. and {Price}, P.~A. and {Rest}, A. and {Saglia}, R.~P. and {Schlafly}, E.~F. and {Smartt}, S.~J. and {Sweeney}, W. and {Wainscoat}, R.~J. and {Burgett}, W.~S. and {Chastel}, S. and {Grav}, T. and {Heasley}, J.~N. and {Hodapp}, K.~W. and {Jedicke}, R. and {Kaiser}, N. and {Kudritzki}, R. -P. and {Luppino}, G.~A. and {Lupton}, R.~H. and {Monet}, D.~G. and {Morgan}, J.~S. and {Onaka}, P.~M. and {Shiao}, B. and {Stubbs}, C.~W. and {Tonry}, J.~L. and {White}, R. and {Ba{\~n}ados}, E. and {Bell}, E.~F. and {Bender}, R. and {Bernard}, E.~J. and {Boegner}, M. and {Boffi}, F. and {Botticella}, M.~T. and {Calamida}, A. and {Casertano}, S. and {Chen}, W. -P. and {Chen}, X. and {Cole}, S. and {Deacon}, N. and {Frenk}, C. and {Fitzsimmons}, A. and {Gezari}, S. and {Gibbs}, V. and {Goessl}, C. and {Goggia}, T. and {Gourgue}, R. and {Goldman}, B. and {Grant}, P. and {Grebel}, E.~K. and {Hambly}, N.~C. and {Hasinger}, G. and {Heavens}, A.~F. and {Heckman}, T.~M. and {Henderson}, R. and {Henning}, T. and {Holman}, M. and {Hopp}, U. and {Ip}, W. -H. and {Isani}, S. and {Jackson}, M. and {Keyes}, C.~D. and {Koekemoer}, A.~M. and {Kotak}, R. and {Le}, D. and {Liska}, D. and {Long}, K.~S. and {Lucey}, J.~R. and {Liu}, M. and {Martin}, N.~F. and {Masci}, G. and {McLean}, B. and {Mindel}, E. and {Misra}, P. and {Morganson}, E. and {Murphy}, D.~N.~A. and {Obaika}, A. and {Narayan}, G. and {Nieto-Santisteban}, M.~A. and {Norberg}, P. and {Peacock}, J.~A. and {Pier}, E.~A. and {Postman}, M. and {Primak}, N. and {Rae}, C. and {Rai}, A. and {Riess}, A. and {Riffeser}, A. and {Rix}, H.~W. and {R{\"o}ser}, S. and {Russel}, R. and {Rutz}, L. and {Schilbach}, E. and {Schultz}, A.~S.~B. and {Scolnic}, D. and {Strolger}, L. and {Szalay}, A. and {Seitz}, S. and {Small}, E. and {Smith}, K.~W. and {Soderblom}, D.~R. and {Taylor}, P. and {Thomson}, R. and {Taylor}, A.~N. and {Thakar}, A.~R. and {Thiel}, J. and {Thilker}, D. and {Unger}, D. and {Urata}, Y. and {Valenti}, J. and {Wagner}, J. and {Walder}, T. and {Walter}, F. and {Watters}, S.~P. and {Werner}, S. and {Wood-Vasey}, W.~M. and {Wyse}, R.},
        title = "{The Pan-STARRS1 Surveys}",
      journal = {arXiv e-prints},
         year = 2016,
        month = dec,
          eid = {arXiv:1612.05560},
        pages = {arXiv:1612.05560},
          doi = {10.48550/arXiv.1612.05560},
archivePrefix = {arXiv},
       eprint = {1612.05560},
 primaryClass = {astro-ph.IM},
       adsurl = {https://ui.adsabs.harvard.edu/abs/2016arXiv161205560C}
}

@article{Magnier_2020,
  title = {Pan-{{STARRS Photometric}} and {{Astrometric Calibration}}},
  author = {Magnier, {\relax Eugene}. A. and Schlafly, {\relax Edward}. F. and Finkbeiner, Douglas P. and Tonry, J. L. and Goldman, B. and R{\"o}ser, S. and Schilbach, E. and Casertano, S. and Chambers, K. C. and Flewelling, H. A. and Huber, M. E. and Price, P. A. and Sweeney, W. E. and Waters, C. Z. and Denneau, L. and Draper, P. W. and Hodapp, K. W. and Jedicke, R. and Kaiser, N. and Kudritzki, R.-P. and Metcalfe, N. and Stubbs, C. W. and Wainscoat, R. J.},
  year = {2020},
  month = nov,
  journal = {The Astrophysical Journal Supplement Series},
  volume = {251},
  number = {1},
  pages = {6},
  issn = {0067-0049, 1538-4365},
  doi = {10.3847/1538-4365/abb82a},
  urldate = {2025-06-26}
}

@article{GaiaCollaboration_2023,
  title = {Gaia {{Data Release}} 3: {{Summary}} of the Content and Survey Properties},
  shorttitle = {{\emph{Gaia}} {{Data Release}} 3},
  author = {{Gaia Collaboration} and Vallenari, A. and Brown, A. G. A. and Prusti, T. and De Bruijne, J. H. J. and Arenou, F. and Babusiaux, C. and Biermann, M. and Creevey, O. L. and Ducourant, C. and Evans, D. W. and Eyer, L. and Guerra, R. and Hutton, A. and Jordi, C. and Klioner, S. A. and Lammers, U. L. and Lindegren, L. and Luri, X. and Mignard, F. and Panem, C. and Pourbaix, D. and Randich, S. and Sartoretti, P. and Soubiran, C. and Tanga, P. and Walton, N. A. and {Bailer-Jones}, C. A. L. and Bastian, U. and Drimmel, R. and Jansen, F. and Katz, D. and Lattanzi, M. G. and Van Leeuwen, F. and Bakker, J. and Cacciari, C. and Casta{\~n}eda, J. and De Angeli, F. and Fabricius, C. and Fouesneau, M. and Fr{\'e}mat, Y. and Galluccio, L. and Guerrier, A. and Heiter, U. and Masana, E. and Messineo, R. and Mowlavi, N. and Nicolas, C. and Nienartowicz, K. and Pailler, F. and Panuzzo, P. and Riclet, F. and Roux, W. and Seabroke, G. M. and Sordo, R. and Th{\'e}venin, F. and {Gracia-Abril}, G. and Portell, J. and Teyssier, D. and Altmann, M. and Andrae, R. and Audard, M. and {Bellas-Velidis}, I. and Benson, K. and Berthier, J. and Blomme, R. and Burgess, P. W. and Busonero, D. and Busso, G. and C{\'a}novas, H. and Carry, B. and Cellino, A. and Cheek, N. and Clementini, G. and Damerdji, Y. and Davidson, M. and De Teodoro, P. and Nu{\~n}ez Campos, M. and Delchambre, L. and Dell'Oro, A. and Esquej, P. and {Fern{\'a}ndez-Hern{\'a}ndez}, J. and Fraile, E. and Garabato, D. and {Garc{\'i}a-Lario}, P. and Gosset, E. and Haigron, R. and Halbwachs, J.-L. and Hambly, N. C. and Harrison, D. L. and Hern{\'a}ndez, J. and Hestroffer, D. and Hodgkin, S. T. and Holl, B. and Jan{\ss}en, K. and Jevardat De Fombelle, G. and Jordan, S. and {Krone-Martins}, A. and Lanzafame, A. C. and L{\"o}ffler, W. and Marchal, O. and Marrese, P. M. and Moitinho, A. and Muinonen, K. and Osborne, P. and Pancino, E. and Pauwels, T. and {Recio-Blanco}, A. and Reyl{\'e}, C. and Riello, M. and Rimoldini, L. and Roegiers, T. and Rybizki, J. and Sarro, L. M. and Siopis, C. and Smith, M. and Sozzetti, A. and Utrilla, E. and Van Leeuwen, M. and Abbas, U. and {\'A}brah{\'a}m, P. and Abreu Aramburu, A. and Aerts, C. and Aguado, J. J. and Ajaj, M. and {Aldea-Montero}, F. and Altavilla, G. and {\'A}lvarez, M. A. and Alves, J. and Anders, F. and Anderson, R. I. and Anglada Varela, E. and Antoja, T. and Baines, D. and Baker, S. G. and {Balaguer-N{\'u}{\~n}ez}, L. and Balbinot, E. and Balog, Z. and Barache, C. and Barbato, D. and Barros, M. and Barstow, M. A. and Bartolom{\'e}, S. and Bassilana, J.-L. and Bauchet, N. and Becciani, U. and Bellazzini, M. and Berihuete, A. and Bernet, M. and Bertone, S. and Bianchi, L. and Binnenfeld, A. and {Blanco-Cuaresma}, S. and Blazere, A. and Boch, T. and Bombrun, A. and Bossini, D. and Bouquillon, S. and Bragaglia, A. and Bramante, L. and Breedt, E. and Bressan, A. and Brouillet, N. and Brugaletta, E. and Bucciarelli, B. and Burlacu, A. and Butkevich, A. G. and Buzzi, R. and Caffau, E. and Cancelliere, R. and {Cantat-Gaudin}, T. and Carballo, R. and Carlucci, T. and Carnerero, M. I. and Carrasco, J. M. and Casamiquela, L. and Castellani, M. and {Castro-Ginard}, A. and Chaoul, L. and Charlot, P. and Chemin, L. and Chiaramida, V. and Chiavassa, A. and Chornay, N. and Comoretto, G. and Contursi, G. and Cooper, W. J. and Cornez, T. and Cowell, S. and Crifo, F. and Cropper, M. and Crosta, M. and Crowley, C. and Dafonte, C. and Dapergolas, A. and David, M. and David, P. and De Laverny, P. and De Luise, F. and De March, R. and De Ridder, J. and De Souza, R. and De Torres, A. and Del Peloso, E. F. and Del Pozo, E. and Delbo, M. and Delgado, A. and Delisle, J.-B. and Demouchy, C. and Dharmawardena, T. E. and Di Matteo, P. and Diakite, S. and Diener, C. and Distefano, E. and Dolding, C. and Edvardsson, B. and Enke, H. and Fabre, C. and Fabrizio, M. and Faigler, S. and Fedorets, G. and Fernique, P. and Fienga, A. and Figueras, F. and Fournier, Y. and Fouron, C. and Fragkoudi, F. and Gai, M. and {Garcia-Gutierrez}, A. and {Garcia-Reinaldos}, M. and {Garc{\'i}a-Torres}, M. and Garofalo, A. and Gavel, A. and Gavras, P. and Gerlach, E. and Geyer, R. and Giacobbe, P. and Gilmore, G. and Girona, S. and Giuffrida, G. and Gomel, R. and Gomez, A. and {Gonz{\'a}lez-N{\'u}{\~n}ez}, J. and {Gonz{\'a}lez-Santamar{\'i}a}, I. and {Gonz{\'a}lez-Vidal}, J. J. and Granvik, M. and Guillout, P. and Guiraud, J. and {Guti{\'e}rrez-S{\'a}nchez}, R. and Guy, L. P. and Hatzidimitriou, D. and Hauser, M. and Haywood, M. and Helmer, A. and Helmi, A. and Sarmiento, M. H. and Hidalgo, S. L. and Hilger, T. and H{\l}adczuk, N. and Hobbs, D. and Holland, G. and Huckle, H. E. and Jardine, K. and Jasniewicz, G. and {Jean-Antoine Piccolo}, A. and {Jim{\'e}nez-Arranz}, {\'O}. and Jorissen, A. and Juaristi Campillo, J. and Julbe, F. and Karbevska, L. and Kervella, P. and Khanna, S. and Kontizas, M. and Kordopatis, G. and Korn, A. J. and K{\'o}sp{\'a}l, {\'A} and {Kostrzewa-Rutkowska}, Z. and Kruszy{\'n}ska, K. and Kun, M. and Laizeau, P. and Lambert, S. and Lanza, A. F. and Lasne, Y. and Le Campion, J.-F. and Lebreton, Y. and Lebzelter, T. and Leccia, S. and Leclerc, N. and {Lecoeur-Taibi}, I. and Liao, S. and Licata, E. L. and Lindstr{\o}m, H. E. P. and Lister, T. A. and Livanou, E. and Lobel, A. and Lorca, A. and Loup, C. and Madrero Pardo, P. and Magdaleno Romeo, A. and Managau, S. and Mann, R. G. and Manteiga, M. and Marchant, J. M. and Marconi, M. and Marcos, J. and Marcos Santos, M. M. S. and Mar{\'i}n Pina, D. and Marinoni, S. and Marocco, F. and Marshall, D. J. and Martin Polo, L. and {Mart{\'i}n-Fleitas}, J. M. and Marton, G. and Mary, N. and Masip, A. and Massari, D. and {Mastrobuono-Battisti}, A. and Mazeh, T. and McMillan, P. J. and Messina, S. and Michalik, D. and Millar, N. R. and Mints, A. and Molina, D. and Molinaro, R. and Moln{\'a}r, L. and Monari, G. and Mongui{\'o}, M. and Montegriffo, P. and Montero, A. and Mor, R. and Mora, A. and Morbidelli, R. and Morel, T. and Morris, D. and Muraveva, T. and Murphy, C. P. and Musella, I. and Nagy, Z. and Noval, L. and Oca{\~n}a, F. and Ogden, A. and Ordenovic, C. and Osinde, J. O. and Pagani, C. and Pagano, I. and Palaversa, L. and Palicio, P. A. and {Pallas-Quintela}, L. and Panahi, A. and {Payne-Wardenaar}, S. and Pe{\~n}alosa Esteller, X. and Penttil{\"a}, A. and Pichon, B. and Piersimoni, A. M. and Pineau, F.-X. and Plachy, E. and Plum, G. and Poggio, E. and Pr{\v s}a, A. and Pulone, L. and Racero, E. and Ragaini, S. and Rainer, M. and Raiteri, C. M. and Rambaux, N. and Ramos, P. and {Ramos-Lerate}, M. and Re Fiorentin, P. and Regibo, S. and Richards, P. J. and Rios Diaz, C. and Ripepi, V. and Riva, A. and Rix, H.-W. and Rixon, G. and Robichon, N. and Robin, A. C. and Robin, C. and Roelens, M. and Rogues, H. R. O. and Rohrbasser, L. and {Romero-G{\'o}mez}, M. and Rowell, N. and Royer, F. and Ruz Mieres, D. and Rybicki, K. A. and Sadowski, G. and S{\'a}ez N{\'u}{\~n}ez, A. and Sagrist{\`a} Sell{\'e}s, A. and Sahlmann, J. and Salguero, E. and Samaras, N. and Sanchez Gimenez, V. and Sanna, N. and Santove{\~n}a, R. and Sarasso, M. and Schultheis, M. and Sciacca, E. and Segol, M. and Segovia, J. C. and S{\'e}gransan, D. and Semeux, D. and Shahaf, S. and Siddiqui, H. I. and Siebert, A. and Siltala, L. and Silvelo, A. and Slezak, E. and Slezak, I. and Smart, R. L. and Snaith, O. N. and Solano, E. and Solitro, F. and Souami, D. and Souchay, J. and Spagna, A. and Spina, L. and Spoto, F. and Steele, I. A. and Steidelm{\"u}ller, H. and Stephenson, C. A. and S{\"u}veges, M. and Surdej, J. and Szabados, L. and {Szegedi-Elek}, E. and Taris, F. and Taylor, M. B. and Teixeira, R. and Tolomei, L. and Tonello, N. and Torra, F. and Torra, J. and Torralba Elipe, G. and Trabucchi, M. and Tsounis, A. T. and Turon, C. and Ulla, A. and Unger, N. and Vaillant, M. V. and Van Dillen, E. and Van Reeven, W. and Vanel, O. and Vecchiato, A. and Viala, Y. and Vicente, D. and Voutsinas, S. and Weiler, M. and Wevers, T. and Wyrzykowski, {\L}. and Yoldas, A. and Yvard, P. and Zhao, H. and Zorec, J. and Zucker, S. and Zwitter, T.},
  year = {2023},
  month = jun,
  journal = {Astronomy \& Astrophysics},
  volume = {674},
  pages = {A1},
  issn = {0004-6361, 1432-0746},
  doi = {10.1051/0004-6361/202243940},
  urldate = {2025-05-29},
  copyright = {https://creativecommons.org/licenses/by/4.0}
}

@INPROCEEDINGS{SkyBot_2006,
       author = {{Berthier}, J. and {Vachier}, F. and {Thuillot}, W. and {Fernique}, P. and {Ochsenbein}, F. and {Genova}, F. and {Lainey}, V. and {Arlot}, J. -E.},
        title = "{SkyBoT, a new VO service to identify Solar System objects}",
    booktitle = {Astronomical Data Analysis Software and Systems XV},
         year = 2006,
       editor = {{Gabriel}, C. and {Arviset}, C. and {Ponz}, D. and {Enrique}, S.},
       series = {Astronomical Society of the Pacific Conference Series},
       volume = {351},
        month = jul,
        pages = {367},
       adsurl = {https://ui.adsabs.harvard.edu/abs/2006ASPC..351..367B}
}

@article{Bellm_2019,
  title = {The {{Zwicky Transient Facility}}: {{System Overview}}, {{Performance}}, and {{First Results}}},
  shorttitle = {The {{Zwicky Transient Facility}}},
  author = {Bellm, Eric C. and Kulkarni, Shrinivas R. and Graham, Matthew J. and Dekany, Richard and Smith, Roger M. and Riddle, Reed and Masci, Frank J. and Helou, George and Prince, Thomas A. and Adams, Scott M. and Barbarino, C. and Barlow, Tom and Bauer, James and Beck, Ron and Belicki, Justin and Biswas, Rahul and Blagorodnova, Nadejda and Bodewits, Dennis and Bolin, Bryce and Brinnel, Valery and Brooke, Tim and Bue, Brian and Bulla, Mattia and Burruss, Rick and Cenko, S. Bradley and Chang, Chan-Kao and Connolly, Andrew and Coughlin, Michael and Cromer, John and Cunningham, Virginia and De, Kishalay and Delacroix, Alex and Desai, Vandana and Duev, Dmitry A. and Eadie, Gwendolyn and Farnham, Tony L. and Feeney, Michael and Feindt, Ulrich and Flynn, David and Franckowiak, Anna and Frederick, S. and Fremling, C. and {Gal-Yam}, Avishay and Gezari, Suvi and Giomi, Matteo and Goldstein, Daniel A. and Golkhou, V. Zach and Goobar, Ariel and Groom, Steven and Hacopians, Eugean and Hale, David and Henning, John and Ho, Anna Y. Q. and Hover, David and Howell, Justin and Hung, Tiara and Huppenkothen, Daniela and Imel, David and Ip, Wing-Huen and Ivezi{\'c}, {\v Z}eljko and Jackson, Edward and Jones, Lynne and Juric, Mario and Kasliwal, Mansi M. and Kaspi, S. and Kaye, Stephen and Kelley, Michael S. P. and Kowalski, Marek and Kramer, Emily and Kupfer, Thomas and Landry, Walter and Laher, Russ R. and Lee, Chien-De and Lin, Hsing Wen and Lin, Zhong-Yi and Lunnan, Ragnhild and Giomi, Matteo and Mahabal, Ashish and Mao, Peter and Miller, Adam A. and Monkewitz, Serge and Murphy, Patrick and Ngeow, Chow-Choong and Nordin, Jakob and Nugent, Peter and Ofek, Eran and Patterson, Maria T. and Penprase, Bryan and Porter, Michael and Rauch, Ludwig and Rebbapragada, Umaa and Reiley, Dan and Rigault, Mickael and Rodriguez, Hector and van Roestel, Jan and Rusholme, Ben and van Santen, Jakob and Schulze, S. and Shupe, David L. and Singer, Leo P. and Soumagnac, Maayane T. and Stein, Robert and Surace, Jason and Sollerman, Jesper and Szkody, Paula and Taddia, F. and Terek, Scott and Van Sistine, Angela and {van Velzen}, Sjoert and Vestrand, W. Thomas and Walters, Richard and Ward, Charlotte and Ye, Quan-Zhi and Yu, Po-Chieh and Yan, Lin and Zolkower, Jeffry},
  year = {2019},
  month = jan,
  journal = {Publications of the Astronomical Society of the Pacific},
  volume = {131},
  number = {995},
  pages = {018002},
  issn = {0004-6280, 1538-3873},
  doi = {10.1088/1538-3873/aaecbe},
  urldate = {2022-05-09}
}

@article{Forster_2021,
  title = {The {{Automatic Learning}} for the {{Rapid Classification}} of {{Events}} ({{ALeRCE}}) {{Alert Broker}}},
  author = {F{\"o}rster, F. and {Cabrera-Vives}, G. and {Castillo-Navarrete}, E. and Est{\'e}vez, P. A. and {S{\'a}nchez-S{\'a}ez}, P. and Arredondo, J. and Bauer, F. E. and {Carrasco-Davis}, R. and Catelan, M. and Elorrieta, F. and Eyheramendy, S. and Huijse, P. and Pignata, G. and Reyes, E. and Reyes, I. and {Rodr{\'i}guez-Mancini}, D. and {Ruz-Mieres}, D. and Valenzuela, C. and {\'A}lvarez-Maldonado, I. and Astorga, N. and Borissova, J. and Clocchiatti, A. and De Cicco, D. and {Donoso-Oliva}, C. and {Hern{\'a}ndez-Garc{\'i}a}, L. and Graham, M. J. and Jord{\'a}n, A. and Kurtev, R. and Mahabal, A. and Maureira, J. C. and {Mu{\~n}oz-Arancibia}, A. and {Molina-Ferreiro}, R. and Moya, A. and Palma, W. and {P{\'e}rez-Carrasco}, M. and Protopapas, P. and Romero, M. and {Sabatini-Gacitua}, L. and S{\'a}nchez, A. and Mart{\'i}n, J. San and {Sep{\'u}lveda-Cobo}, C. and Vera, E. and Vergara, J. R.},
  year = {2021},
  month = may,
  journal = {The Astronomical Journal},
  volume = {161},
  number = {5},
  pages = {242},
  issn = {0004-6256, 1538-3881},
  doi = {10.3847/1538-3881/abe9bc},
  urldate = {2025-06-30}
}

@article{graham_zwicky_2019,
	title = {The {Zwicky} {Transient} {Facility}: {Science} {Objectives}},
	volume = {131},
	issn = {0004-6280, 1538-3873},
	shorttitle = {The {Zwicky} {Transient} {Facility}},
	url = {https://iopscience.iop.org/article/10.1088/1538-3873/ab006c},
	doi = {10.1088/1538-3873/ab006c},
	number = {1001},
	urldate = {2022-05-09},
	journal = {Publications of the Astronomical Society of the Pacific},
	author = {Graham, Matthew J. and Kulkarni, S. R. and Bellm, Eric C. and Adams, Scott M. and Barbarino, Cristina and Blagorodnova, Nadejda and Bodewits, Dennis and Bolin, Bryce and Brady, Patrick R. and Cenko, S. Bradley and Chang, Chan-Kao and Coughlin, Michael W. and De, Kishalay and Eadie, Gwendolyn and Farnham, Tony L. and Feindt, Ulrich and Franckowiak, Anna and Fremling, Christoffer and Gezari, Suvi and Ghosh, Shaon and Goldstein, Daniel A. and Golkhou, V. Zach and Goobar, Ariel and Ho, Anna Y. Q. and Huppenkothen, Daniela and Ivezić, Željko and Jones, R. Lynne and Juric, Mario and Kaplan, David L. and Kasliwal, Mansi M. and Kelley, Michael S. P. and Kupfer, Thomas and Lee, Chien-De and Lin, Hsing Wen and Lunnan, Ragnhild and Mahabal, Ashish A. and Miller, Adam A. and Ngeow, Chow-Choong and Nugent, Peter and Ofek, Eran O. and Prince, Thomas A. and Rauch, Ludwig and Roestel, Jan van and Schulze, Steve and Singer, Leo P. and Sollerman, Jesper and Taddia, Francesco and Yan, Lin and Ye, Quan-Zhi and Yu, Po-Chieh and Barlow, Tom and Bauer, James and Beck, Ron and Belicki, Justin and Biswas, Rahul and Brinnel, Valery and Brooke, Tim and Bue, Brian and Bulla, Mattia and Burruss, Rick and Connolly, Andrew and Cromer, John and Cunningham, Virginia and Dekany, Richard and Delacroix, Alex and Desai, Vandana and Duev, Dmitry A. and Feeney, Michael and Flynn, David and Frederick, Sara and Gal-Yam, Avishay and Giomi, Matteo and Groom, Steven and Hacopians, Eugean and Hale, David and Helou, George and Henning, John and Hover, David and Hillenbrand, Lynne A. and Howell, Justin and Hung, Tiara and Imel, David and Ip, Wing-Huen and Jackson, Edward and Kaspi, Shai and Kaye, Stephen and Kowalski, Marek and Kramer, Emily and Kuhn, Michael and Landry, Walter and Laher, Russ R. and Mao, Peter and Masci, Frank J. and Monkewitz, Serge and Murphy, Patrick and Nordin, Jakob and Patterson, Maria T. and Penprase, Bryan and Porter, Michael and Rebbapragada, Umaa and Reiley, Dan and Riddle, Reed and Rigault, Mickael and Rodriguez, Hector and Rusholme, Ben and Santen, Jakob van and Shupe, David L. and Smith, Roger M. and Soumagnac, Maayane T. and Stein, Robert and Surace, Jason and Szkody, Paula and Terek, Scott and Sistine, Angela Van and Velzen, Sjoert van and Vestrand, W. Thomas and Walters, Richard and Ward, Charlotte and Zhang, Chaoran and Zolkower, Jeffry},
	month = jul,
	year = {2019},
	pages = {078001},
}

@ARTICLE{ATLAS_force,
       author = {{Shingles}, L. and {Smith}, K.~W. and {Young}, D.~R. and {Smartt}, S.~J. and {Tonry}, J. and {Denneau}, L. and {Heinze}, A. and {Weiland}, H. and {Flewelling}, H. and {Stalder}, B. and {Clocchiatti}, A. and {F{\"o}rster}, F. and {Pignata}, G. and {Rest}, A. and {Anderson}, J. and {Stubbs}, C. and {Erasmus}, N.},
        title = "{Release of the ATLAS Forced Photometry server for public use}",
      journal = {Transient Name Server AstroNote},
         year = 2021,
        month = jan,
       volume = {7},
        pages = {1-7},
       adsurl = {https://ui.adsabs.harvard.edu/abs/2021TNSAN...7....1S}
}

@article{Smith_2020,
  title = {Design and {{Operation}} of the {{ATLAS Transient Science Server}}},
  author = {Smith, K. W. and Smartt, S. J. and Young, D. R. and Tonry, J. L. and Denneau, L. and Flewelling, H. and Heinze, A. N. and Weiland, H. J. and Stalder, B. and Rest, A. and Stubbs, C. W. and Anderson, J. P. and Chen, T.-W and Clark, P. and Do, A. and F{\"o}rster, F. and Fulton, M. and Gillanders, J. and McBrien, O. R. and O'Neill, D. and Srivastav, S. and Wright, D. E.},
  year = {2020},
  month = aug,
  journal = {Publications of the Astronomical Society of the Pacific},
  volume = {132},
  number = {1014},
  pages = {085002},
  issn = {0004-6280, 1538-3873},
  doi = {10.1088/1538-3873/ab936e},
  urldate = {2025-06-30}
}

@article{Tonry_2018,
  title = {{{ATLAS}}: {{A High-cadence All-sky Survey System}}},
  shorttitle = {{{ATLAS}}},
  author = {Tonry, J. L. and Denneau, L. and Heinze, A. N. and Stalder, B. and Smith, K. W. and Smartt, S. J. and Stubbs, C. W. and Weiland, H. J. and Rest, A.},
  year = {2018},
  month = jun,
  journal = {Publications of the Astronomical Society of the Pacific},
  volume = {130},
  number = {988},
  pages = {064505},
  issn = {0004-6280, 1538-3873},
  doi = {10.1088/1538-3873/aabadf},
  urldate = {2025-06-30}
}

@INPROCEEDINGS{TNS,
       author = {{Gal-Yam}, A.},
        title = "{The TNS alert system}",
    booktitle = {American Astronomical Society Meeting Abstracts},
         year = 2021,
       series = {American Astronomical Society Meeting Abstracts},
       volume = {237},
        month = jan,
          eid = {423.05},
        pages = {423.05},
       adsurl = {https://ui.adsabs.harvard.edu/abs/2021AAS...23742305G}
}

@misc{desicollaboration2025datarelease1dark,
      title={Data Release 1 of the Dark Energy Spectroscopic Instrument}, 
      author={DESI Collaboration and M. Abdul-Karim and A. G. Adame and D. Aguado and J. Aguilar and S. Ahlen and S. Alam and G. Aldering and D. M. Alexander and R. Alfarsy and L. Allen and C. Allende Prieto and O. Alves and A. Anand and U. Andrade and E. Armengaud and S. Avila and A. Aviles and H. Awan and S. Bailey and A. Baleato Lizancos and O. Ballester and A. Bault and J. Bautista and S. BenZvi and L. Beraldo e Silva and J. R. Bermejo-Climent and F. Beutler and D. Bianchi and C. Blake and R. Blum and A. S. Bolton and M. Bonici and S. Brieden and A. Brodzeller and D. Brooks and E. Buckley-Geer and E. Burtin and R. Canning and A. Carnero Rosell and A. Carr and P. Carrilho and L. Casas and F. J. Castander and R. Cereskaite and J. L. Cervantes-Cota and E. Chaussidon and J. Chaves-Montero and S. Chen and X. Chen and T. Claybaugh and S. Cole and A. P. Cooper and M. -C. Cousinou and A. Cuceu and T. M. Davis and K. S. Dawson and R. de Belsunce and R. de la Cruz and A. de la Macorra and A. de Mattia and N. Deiosso and J. Della Costa and R. Demina and U. Demirbozan and J. DeRose and A. Dey and B. Dey and J. Ding and Z. Ding and P. Doel and K. Douglass and M. Dowicz and H. Ebina and J. Edelstein and D. J. Eisenstein and W. Elbers and N. Emas and S. Escoffier and P. Fagrelius and X. Fan and K. Fanning and V. A. Fawcett and E. Fernández-García and S. Ferraro and N. Findlay and A. Font-Ribera and J. E. Forero-Romero and D. Forero-Sánchez and C. S. Frenk and B. T. Gänsicke and L. Galbany and J. García-Bellido and C. Garcia-Quintero and L. H. Garrison and E. Gaztañaga and H. Gil-Marín and O. Y. Gnedin and S. Gontcho A Gontcho and A. X. Gonzalez-Morales and V. Gonzalez-Perez and C. Gordon and O. Graur and D. Green and D. Gruen and R. Gsponer and C. Guandalin and G. Gutierrez and J. Guy and C. Hahn and J. J. Han and J. Han and S. He and H. K. Herrera-Alcantar and K. Honscheid and J. Hou and C. Howlett and D. Huterer and V. Iršič and M. Ishak and A. Jacques and J. Jimenez and Y. P. Jing and B. Joachimi and S. Joudaki and R. Joyce and E. Jullo and S. Juneau and N. G. Karaçaylı and T. Karim and R. Kehoe and S. Kent and A. Khederlarian and D. Kirkby and T. Kisner and F. -S. Kitaura and N. Kizhuprakkat and H. Kong and S. E. Koposov and A. Kremin and A. Krolewski and O. Lahav and Y. Lai and C. Lamman and T. -W. Lan and M. Landriau and D. Lang and J. U. Lange and J. Lasker and J. M. Le Goff and L. Le Guillou and A. Leauthaud and M. E. Levi and S. Li and T. S. Li and K. Lodha and M. Lokken and Y. Luo and C. Magneville and M. Manera and C. J. Manser and D. Margala and P. Martini and M. Maus and J. McCullough and P. McDonald and G. E. Medina and L. Medina-Varela and A. Meisner and J. Mena-Fernández and A. Menegas and M. Mezcua and R. Miquel and P. Montero-Camacho and J. Moon and J. Moustakas and A. Muñoz-Gutiérrez and D. Muñoz-Santos and A. D. Myers and J. Myles and S. Nadathur and J. Najita and L. Napolitano and J. A. Newman and F. Nikakhtar and R. Nikutta and G. Niz and H. E. Noriega and N. Padmanabhan and E. Paillas and N. Palanque-Delabrouille and A. Palmese and J. Pan and Z. Pan and D. Parkinson and J. Peacock and W. J. Percival and A. Pérez-Fernández and I. Pérez-Ràfols and P. Peterson and J. Piat and M. M. Pieri and M. Pinon and C. Poppett and A. Porredon and F. Prada and R. Pucha and F. Qin and D. Rabinowitz and A. Raichoor and C. Ramírez-Pérez and S. Ramirez-Solano and M. Rashkovetskyi and C. Ravoux and A. H. Riley and A. Rocher and C. Rockosi and J. Rohlf and A. J. Ross and G. Rossi and R. Ruggeri and V. Ruhlmann-Kleider and C. G. Sabiu and K. Said and A. Saintonge and L. Samushia and E. Sanchez and N. Sanders and C. Saulder and E. F. Schlafly and D. Schlegel and D. Scholte and M. Schubnell and H. Seo and A. Shafieloo and R. Sharples and J. Silber and M. Siudek and A. Smith and D. Sprayberry and J. Suárez-Pérez and J. Swanson and T. Tan and G. Tarlé and P. Taylor and G. Thomas and R. Tojeiro and R. J. Turner and W. Turner and L. A. Ureña-López and R. Vaisakh and M. Valluri and M. Vargas-Magaña and L. Verde and M. Walther and B. Wang and M. S. Wang and W. Wang and B. A. Weaver and N. Weaverdyck and R. H. Wechsler and M. White and M. Wolfson and J. Yang and C. Yèche and S. Youles and J. Yu and S. Yuan and E. A. Zaborowski and P. Zarrouk and H. Zhang and C. Zhao and R. Zhao and Z. Zheng and R. Zhou and H. Zou and S. Zou and Y. Zu},
      year={2025},
      eprint={2503.14745},
      archivePrefix={arXiv},
      primaryClass={astro-ph.CO},
      url={https://arxiv.org/abs/2503.14745}, 
}

@article{Dalya_2022,
  title = {{{GLADE}}+ : An Extended Galaxy Catalogue for Multimessenger Searches with Advanced Gravitational-Wave Detectors},
  shorttitle = {{{GLADE}}+},
  author = {D{\'a}lya, G and D{\'i}az, R and Bouchet, F R and Frei, Z and Jasche, J and Lavaux, G and Macas, R and Mukherjee, S and P{\'a}lfi, M and {de~Souza}, R S and Wandelt, B D and Bilicki, M and Raffai, P},
  year = {2022},
  month = jun,
  journal = {Monthly Notices of the Royal Astronomical Society},
  volume = {514},
  number = {1},
  pages = {1403--1411},
  issn = {0035-8711, 1365-2966},
  doi = {10.1093/mnras/stac1443},
  urldate = {2025-05-08},
  copyright = {https://creativecommons.org/licenses/by/4.0/}
}

@article{WISE-PS1-STRM,
    author = {Kunsági-Máté, Sándor and Beck, Róbert and Szapudi, István and Csabai, István},
    title = {Photometric redshifts for quasars from WISE-PS1-STRM},
    journal = {Monthly Notices of the Royal Astronomical Society},
    volume = {516},
    number = {2},
    pages = {2662-2670},
    year = {2022},
    month = {10},
    issn = {0035-8711},
    doi = {10.1093/mnras/stac2411},
    url = {https://doi.org/10.1093/mnras/stac2411},
    eprint = {https://academic.oup.com/mnras/article-pdf/516/2/2662/45809367/stac2411.pdf},
}

@article{Pastorello_2019a,
  title = {Supernova Impostors and Other Gap Transients},
  author = {Pastorello, Andrea and Fraser, Morgan},
  year = 2019,
  month = aug,
  journal = {Nature Astronomy},
  volume = {3},
  number = {8},
  pages = {676--679},
  issn = {2397-3366},
  doi = {10.1038/s41550-019-0809-9},
  urldate = {2025-08-14}
}
\bibliographystyle{aasjournal}



\end{document}